\documentclass[sigconf]{acmart}

\AtBeginDocument{%
  }

\usepackage{multirow}
\copyrightyear{2025}
\acmYear{2025}
\setcopyright{acmlicensed}\acmConference[CHI '25]{CHI Conference on Human Factors in Computing Systems}{April 26-May 1, 2025}{Yokohama, Japan}
\acmBooktitle{CHI Conference on Human Factors in Computing Systems (CHI '25), April 26-May 1, 2025, Yokohama, Japan}
\acmDOI{10.1145/3706598.3713360}
\acmISBN{979-8-4007-1394-1/2025/04}

\begin{document}

\title[\small{YouthCare}]{YouthCare: Building a Personalized Collaborative Video Censorship Tool to Support Parent-Child Joint Media Engagement}

\author{Wenxin Zhao}
\affiliation{%
  \institution{Fudan University}
  \city{Shanghai}
  \country{China}}
\email{zhaowx21@m.fudan.edu.cn}

\author{Fangyu Yu}
\affiliation{%
  \institution{Fudan University}
  \city{Shanghai}
  \country{China}}
\email{yufy16@fudan.edu.cn}

\author{Peng Zhang}
\authornote{Corresponding authors.}
\affiliation{%
  \institution{Fudan University}
  \city{Shanghai}
  \country{China}}
\email{zhangpeng_@fudan.edu.cn}

\author{Hansu Gu}
\affiliation{%
  \institution{Independent}
  \city{Seattle}
  \country{Washington, USA}}
\email{hansug@acm.org}

\author{Lin Wang}
\affiliation{%
  \institution{Fudan University}
  \city{Shanghai}
  \country{China}}
\email{linw@fudan.edu.cn}

\author{Siyuan Qiao}
\affiliation{%
  \institution{Fudan University}
  \city{Shanghai}
  \country{China}}
\email{syqiao23@m.fudan.edu.cn}

\author{Tun Lu}
\authornotemark[1]
\affiliation{%
  \institution{Fudan University}
  \city{Shanghai}
  \country{China}}
\email{lutun@fudan.edu.cn}

\author{Ning Gu}
\affiliation{%
  \institution{Fudan University}
  \city{Shanghai}
  \country{China}}
\email{ninggu@fudan.edu.cn}

\renewcommand{\shortauthors}{Zhao et al.}

\begin{abstract}
To mitigate the negative impacts of online videos on teenagers, existing research and platforms have implemented various parental mediation mechanisms, such as Parent-Child Joint Media Engagement (JME). However, JME generally relies heavily on parents' time, knowledge, and experience. To fill this gap, we aim to design an automatic tool to help parents/children censor videos more effectively and efficiently in JME. For this goal, we first conducted a formative study to identify the needs and expectations of teenagers and parents for such a system. Based on the findings, we designed YouthCare, a personalized collaborative video censorship tool that supports parents and children to collaboratively filter out inappropriate content and select appropriate content in JME. An evaluation with 10 parent-child pairs demonstrated YouthCare's several strengths in supporting video censorship, while also highlighting some potential problems. These findings inspire us to propose several insights for the future design of parent-child collaborative JME systems.


\end{abstract}



\maketitle

\section{Introduction}
\label{Introduction}
With the advancement of the digital era, more and more teenagers are increasingly engaging with the Internet. Currently, one-third of the global Internet users are teenagers, with usage rates among 12-17 years old exceeding 98\% in Canada and over 96\% in the U.S. \cite{canada_statistical, us_statistical}. In China, the number of adolescent Internet users has increased from 169 million to 193 million over the past five years, with the penetration rate among teenagers increasing from 93.7\% to 97.2\%, and reaching 95.1\% among primary school students \cite{Report2023}. In this context, video platforms have become highly favored by teenagers as one of the primary resources for online learning and entertainment \cite{video_platform}. For instance, a survey on teenagers' Internet usage revealed that about half of teenagers (47.5\%) choose to watch videos to satisfy their needs for knowledge exploration, learning, and entertainment \cite{thomas2000comparing}. However, existing video platforms often lack design considerations specific to teenagers, and there is a significant mismatch between platform content and the characteristics of this age group. As a result, teenagers are often exposed to unhealthy or inappropriate video content, such as violence, pornography, and crime \cite{yu2024parent}. This severely impacts their online experience and even physical and mental health. According to Piaget's theory of cognitive development, teenagers, particularly those in the concrete operational stage (7-11 years old) and the formal operational stage (11 years and older), have limitations in self-cognition, abstract thinking, and experiential reasoning \cite{huitt2003piaget}. For teenagers in these stages, parental involvement is crucial for their growth and development \cite{brown2015beyond}. Such involvement not only helps protect teenagers from potential harms but also fosters intergenerational communication, enhances mutual understanding and trust, and cultivates a positive family environment \cite{nathanson2000reducing}.

In light of this, existing research and video platforms have introduced various parental mediation mechanisms to mitigate the negative impacts of the online video environment on teenagers \cite{nathanson2000reducing}. 
For example, the Parental Mediation Theory (PMT) proposes three strategies for parents to help their children maximize the benefits and minimize the risks associated with media use: restrictive mediation, active mediation, and co-using \cite{jiow2017level}.
The Joint Media Engagement (JME) framework \cite{yu2024parent} emphasizes the necessity of joint participation between parents and children. Recently, there has also been growing advocacy for parents to use media technology together with their children \cite{connell2015parental}. 
For instance, the American Academy of Pediatrics (AAP) supports this approach in its policies, encouraging parents to shift from limiting children's screen time to co-using media with their children \cite{reid2016children}. This transformation helps foster children's learning and social development \cite{griffith2022making} and strengthen family relationships \cite{musick2021gaming} through shared experiences. 
However, parental mediation, especially co-using, relies heavily on parents' time, knowledge, and experience. On the one hand, parents need to select appropriate video content for their children and understand the various settings tailored for teenagers, which is cumbersome, time-consuming, and labor-intensive. On the other hand, due to limitations in knowledge and experience, the content filtered by parents for their children may still be exposed to risks and inconsistent with teenagers' interests.

The development of multimodal large language models (MLLMs) offers a promising solution to the above issues. Firstly, MLLMs show strong capabilities in parsing and understanding text, video, and other forms of media. Pre-trained on large-scale multimodal corpora by integrating natural language processing (NLP) and computer vision technologies, MLLMs can effectively analyze and understand textual content associated with online videos, such as titles and tags, as well as the video content itself. These models have been widely applied in fields such as content moderation \cite{zellers2019defending} and toxicity detection \cite{schmidt2017survey}. Secondly, MLLMs exhibit remarkable interactive capabilities through instruction learning \cite{liu2024visual}, in-context learning \cite{zhao2023mmicl}, and few-shot learning \cite{tsimpoukelli2021multimodal}. This enables MLLMs to quickly grasp user intentions and generate responses, making them broadly applicable in areas such as emotional support \cite{bilquise2022emotionally} and intelligent customer service \cite{almansor2020survey}. Lastly, MLLMs are easy to use without additional training requirements, which makes them easy to deploy in real-world scenarios \cite{chen2022simple}.

The background above inspires us to explore whether it is possible to design a video censorship tool based on MLLMs to help parents/children select appropriate videos more effectively and efficiently in JME. What needs to be clarified is that the objects considered in this work are parents/children who are aware of risk videos and willing to take some actions. Individuals who are unaware of the risks or unwilling to take actions are not within the scope of this study. 
However, designing a video censorship tool based on MLLMs to support JME faces several challenges. Firstly, the expectations and needs of teenagers and their parents regarding such a tool are unclear. For example, how to alleviate parental burden while ensuring their participation, what roles the MLLMs should play in this process, and how to effectively promote collaboration between the system, parents, and children, all of which hinder the design of this tool. Secondly, assisting teenagers and parents in censoring video content may involve multiple potential tasks, such as effectively filtering out risky content, identifying the content of interest, and facilitating collaboration between teenagers and parents. Jointly achieving these goals with large language models is complicated. Lastly, to validate the effectiveness of the tool, it is necessary to recruit some pairs of parents and teenagers to use and evaluate it in real-world scenarios. Some metrics, such as the appropriateness of selected videos, are difficult to assess, as they often rely on personalized judgment and context, aggravating the complexity of this research.

Considering these issues and challenges, our aim is to design an automated video censorship tool to facilitate collaborative filtering of inappropriate content and selection of appropriate content by parents and children in JME. Our study focuses on teenagers aged 8-15. This age group possesses basic social media skills and can perform fundamental tasks such as watching videos and adjusting settings \cite{aarts2022snoozy}. However, they still face limitations in self-cognition, abstract thinking, and experiential reasoning, necessitating special attention during video consumption \cite{huitt2003piaget}. These characteristics make teenagers within this age range suitable subjects for JME research. We began by conducting a formative study with these teenagers and their parents. Through questionnaires and semi-structured interviews, we explored their needs and expectations for a video censorship tool. Based on our findings, we identified five design goals: conducting video censorship from both risk and suitability perspectives; supporting collaborative configuration of personalized censorship needs through mediation; supporting both direct and indirect methods for configuring personalized censorship needs; providing both in-time and long-term explanatory feedback on censorship needs; and ensuring user control. Guided by these goals, we designed YouthCare, a personalized collaborative video censorship tool to help parents and children collaboratively filter out inappropriate content and select appropriate content in JME. YouthCare comprises three main modules: Collaborative Configuration Module, Personalized Video Censorship Module, and Feedback Report Module. The Collaborative Configuration Module enables parents and teenagers to jointly configure censorship objects directly (using keywords) or indirectly (using representative videos). It integrates a Chatbot-based consensus-building component to help resolve conflicts in censorship objects between child and parent, generating a ``co-preference'' for video censorship. Building on this, the Personalized Video Censorship Module offers video censorship and rationale explanation, considering both Personalized Guidelines (derived from co-preference) and Common Guidelines (based on authoritative standards and policies). Following that, the Feedback Report Module delivers in-time feedback and long-term summarization of detection results, tailored to align with the established co-preference. All modules are implemented using MLLMs. We recruited 10 pairs of children and parents to use and evaluate this tool. Results indicated that YouthCare performs well in meeting user needs and demonstrates high acceptance and willingness for continued use. Based on these findings, we provide new insights into the design of future parent-child collaborative JME systems. To conclude, the contributions of this paper can be summarized as follows:
\begin{itemize}
    \item To the best of our knowledge, this is the first work of leveraging MLLMs to help children and parents collaboratively moderate video content in JME.
    
    \item We conduct a formative study that reveals the needs and expectations of parents and teenagers regarding the design of video censorship tools.
    
    \item We design and implement YouthCare to help parents and children collaboratively filter out inappropriate content and select appropriate content in JME. Extensive evaluations are also conducted to validate its performance and user experience.  
    
    \item We propose several new insights into designing parent-child collaborative JME systems.
\end{itemize}

The remainder of this paper is organized as follows. First, we present related work to introduce the context of the current research in Section~\ref{Related Works}. In Section~\ref{Formative Study}, we describe the procedure and results of our formative study. Section~\ref{YouthCare} details the framework of YouthCare and its implementation. The evaluation methods and corresponding results are presented in Section~\ref{Evaluation}. Section~\ref{Discussion and Implications} and Section~\ref{Ethical Consideration} discuss the findings and the ethical consideration of our work, respectively. And Section~\ref{Limitations and Future Works} indicates our limitations and future works. Finally, we present our conclusions in Section~\ref{Conclusion}.

\section{Related Works}
\label{Related Works}

\subsection{Parental Mediation Theory and Joint Media Engagement}

Parental Mediation Theory (PMT) explains how parents mediate their children's media behaviors through various strategies, aiming to maximize benefits and minimize risks in children's media use \cite{yu2024parent}.
It includes three mediation strategies: restrictive mediation, active mediation, and co-using \cite{yu2021parental}.
Restrictive mediation emphasizes parental control by setting rules on content, usage times, and duration \cite{hiniker2016not} or directly switching to Youth Mode \cite{qu2023research}.
Active mediation indicates that parents discuss media content and its use with their children, such as content genres and the reliability of websites \cite{eastin2006parenting}, or provide guidance on the usage of the platforms and devices through conversation \cite{ jiow2017level}.
Co-using implies that parents engage in media activities with children to enhance learning and enjoyment by pre-selecting appropriate content and consuming it together \cite{singh2019kidsguard}. 
As parental mediation becomes more prevalent, all three types of mediation strategies are widely used. Among these, the co-using strategy has gained increasing attention for its advantages in fostering stronger parent-child relationships and promoting media engagement in more collaborative ways \cite{connell2015parental}.


A framework closely related to parental mediation, and co-using in particular, is the Joint Media Engagement (JME), which is defined as ``spontaneous and designed experiences of people using media together'' \cite{takeuchi2011new}. It highlights shared media experiences among multiple participants, including parents and children, friends, and others \cite{wang2024koala}. JME, extensively studied in Human-Computer Interaction (HCI), focuses on the parent-child activities, such as co-viewing \cite{sundqvist2021digital} and co-designing \cite{ballagas2013electric}, which helps HCI researchers better understand parent-child JME relationships and develop new technologies for families. Studies demonstrate that these activities can enhance parents' and children's mutual understanding and media practices, significantly improving family relationships \cite{yu2023family}.

However, existing studies on parent-child JME often assume that parents have sufficient time, knowledge, and experience to actively engage in and guide children's media use. However, parents frequently face constraints like work pressure and limited knowledge, hindering their participation. 
Our work addresses this issue by designing an automated and interactive tool to reduce the burden on parents during the JME process. This tool does not rely heavily on parents' time and expertise, thereby promoting broader family engagement and more effective parent-child media collaboration.

\subsection{Child-Parent-Agent Interaction}

In many families, interactive systems - leveraging techniques such as natural language processing, computer vision, and machine learning - have become essential for facilitating communication and interaction between children and parents \cite{sarker2022ai}.
Research suggests that these systems can serve as intelligent agents, enhancing learning and entertainment while strengthening parent-child interactions \cite{novianti2020parental}. These systems are also reported to offer personalized educational and entertainment content by interacting with users through real-time text, voice, and video guidance \cite{missaouib2021interactive}. 

Many studies have focused on children's use of conversational agents for collaborative interaction in activities such as video viewing \cite{du2021alexa, xu2020elinor} and storytelling \cite{yuan2022wordcraft}. In video viewing contexts, \cite{du2021alexa} examines child-agent communication to identify design considerations for children. The study finds that the agent can support communication and prevent breakdowns by repeating answers and varying pitch during voice interactions. Similarly, \cite{xu2020elinor} designs a conversational agent to facilitate interaction between children and an on-screen character, optimizing discussion prompts and feedback to avoid communication breakdowns. It is revealed that this agent tailors conversations to align with children's developmental needs, and thus enhance their engagement during video viewing and learning. Parents can also be involved in the interactive agent to help ensure children's online safety and promote child-parent communication \cite{yu2024parent}. Research indicates that parents are increasingly participating in media use with their children, and many policies are beginning to support the co-using of media \cite{connell2015parental}. By leveraging agents, parents can gain better insights into children's interests and needs, while agents also provide channels for children to express their preferences, thereby fostering communication and interaction between parents and children \cite{zhang2022storybuddy}. For instance, StoryBuddy integrates parent-agent co-reading with child-agent interactions to enhance children's engagement \cite{zhang2022storybuddy}. It helps improve children's reading comprehension and fosters more meaningful interactions between parents and children.

However, prior research on child-parent-agent interaction in video viewing mainly focuses on individual participation of teenagers or relies heavily on pre-configured parental settings, lacking support for collaborative involvement between parents and children. The systems in these activities often lack effective approaches to support joint participation between parents and children in video selection and viewing. This makes it difficult for parents to guide and educate their children about video content while also limiting the children's ability to express their preferences. 
Our research fills this gap by developing an automated video censorship tool to enhance collaborative participation between parents and teenagers, and thus better meet family needs in education and entertainment.

\subsection{Large Language Models in Video Understanding}

Video understanding has evolved from basic classification and recognition to more complex tasks that require logical and commonsense reasoning \cite{tang2023video}. This evolution requires the models to approach human levels of video comprehension. Traditional video understanding models rely on handcrafted feature extraction techniques to capture key information in videos, such as Scale-Invariant Feature Transform (SIFT) \cite{lindeberg2012scale} and Speeded-Up Robust Features (SURF) \cite{bay2008speeded}. Later, deep learning methods that adopt advanced neural network architectures have significantly improved the performance of video understanding. For instance, Convolutional Neural Networks (CNNs) \cite{karpathy2014large} achieve good performance by processing inputs at two spatial resolutions: a low-resolution context stream and a high-resolution fovea stream. Temporal Segment Networks (TSN) \cite{wang2016temporal} enhance the performance of long videos by using a sparse sampling scheme to extract short snippets, which are then uniformly distributed along the temporal dimension and aggregated through a segmental structure.

Recently, Large Language Models (LLMs) have exhibited exceptional capabilities in video understanding, particularly in open-ended spatio-temporal reasoning that integrates commonsense knowledge \cite{tang2023video}. These models benefit from extensive pre-training on large datasets, allowing them to perform various tasks with minimal fine-tuning \cite{zhang2023dnagpt}. Multimodal large language models (MLLMs) extend the capabilities of LLMs by integrating and processing diverse data modalities, such as video, image, and text. This integration makes MLLMs more versatile task solvers, adept at handling a broader range of complex tasks \cite{tang2023video}. In HCI, MLLMs have been demonstrated to enhance the analysis of instructional videos in interactive environments \cite{gan2023large}, support communication for the deaf and hard of hearing through sign language translation \cite{liu2023survey}, and improve user experiences through interactive games and virtual environments \cite{gokce2023role}. In the social media domain, MLLMs have been validated as effective in improving video understanding through subtitle generation \cite{yang2023vid2seq}, content summarization \cite{pramanick2023egovlpv2}, and other approaches.

Although existing research has made achievements in video understanding, it focuses primarily on generalized approaches and lacks fine-grained human interaction involvement \cite{tang2023video}. This limitation is particularly evident when addressing the personalized video comprehension needs and preferences for specific user groups (e.g., teenagers) in specific contexts (e.g., family education). Our work addresses these gaps by integrating multidimensional fine-grained video understanding with personalized user engagement. This approach helps to offer tailored solutions that better meet the needs of parents and children in family settings.

\section{Formative Study}
\label{Formative Study}
To inform our design process, we conducted a formative study to investigate current practices of parental involvement in teenagers' media use. This study aimed to uncover existing challenges and explore potential solutions.

\subsection{Method}
\subsubsection{Participants}
\label{Participants}

We recruited participants through online and offline surveys in primary and middle schools. The inclusion criteria were: (1) each pair consisted of one teenager aged 8 to 15 and their parent (either father or mother); (2) the teenager spent at least 30 minutes per week on video platforms and demonstrated the ability to independently perform basic media interactions, such as searching and selecting video categories; (3) the pairs were aware of the risks associated with online videos and expressed a willingness to mitigate these risks. We selected 11 parent-child pairs for the study. Table~\ref{tab:Summary of participants} provides detailed participant information, where YP represents the parent-child pair (Y denotes the Youth, and P denotes the Parent).
The Parent Usage and Youth Usage refer to the use of video platforms by a parent and a child, respectively. Notably, most parents among our participants are aged between 25 and 44 and have a bachelor's degree or higher. It suggests that parents with higher education background are more likely to consider children's online video-related risks, which is consistent with existing research that highlights a correlation between parents' level of education and their involvement in children's media regulation \cite{livingstone2015parents, cingel2013predicting}.



\begin{table*}[ht] 
\small
\caption{Demographics of formative study participants.}
\label{tab:Summary of participants}
\centering
\resizebox{\linewidth}{!}{
\begin{tabular}{c|ccccccccc}
\hline
\textbf{ID}   & \textbf{\begin{tabular}[c]{@{}c@{}}Youth\\ Age\end{tabular}} & \textbf{\begin{tabular}[c]{@{}c@{}}Youth\\ Gender\end{tabular}} & \textbf{\begin{tabular}[c]{@{}c@{}}Parent\\ Age\end{tabular}} &
\textbf{\begin{tabular}[c]{@{}c@{}}Parent\\ Gender\end{tabular}}&
\textbf{\begin{tabular}[c]{@{}c@{}}Parent\\ Education\end{tabular}}&
\textbf{\begin{tabular}[c]{@{}c@{}}Parent\\ Occupation\end{tabular}}&
\textbf{\begin{tabular}[c]{@{}c@{}}Parent Usage\\ (Years/Frequency)\end{tabular}}&
\textbf{\begin{tabular}[c]{@{}c@{}}Youth Usage\\ (Years/Frequency)\end{tabular}}&
\textbf{\begin{tabular}[c]{@{}c@{}}Types of\\ Mediation\end{tabular}}         \\ 
\midrule
\textbf{YP1}  & 8                                         & Girl            & 35-44                              & Male      &  \begin{tabular}[c]{@{}c@{}}Bachelor's\\ Degree\end{tabular}  &  \begin{tabular}[c]{@{}c@{}}Company\\ Employee\end{tabular} & \begin{tabular}[c]{@{}c@{}}7 Years or More/\\ Every Day\end{tabular}                             & \begin{tabular}[c]{@{}c@{}}Less than 1 Year/\\ 1-6 Days Per Week\end{tabular} & \begin{tabular}[c]{@{}c@{}}Co-using,\\ Restrictive Mediation\end{tabular}\\
\midrule
\textbf{YP2}  & 9                                         & Girl            & 25-34                              & Female    & \begin{tabular}[c]{@{}c@{}}Associate's\\ Degree or Below\end{tabular}   &  Teacher& \begin{tabular}[c]{@{}c@{}}3-5 Years/\\ Every Day\end{tabular}                             & \begin{tabular}[c]{@{}c@{}}1-3 Years/\\ 1-6 Days Per Week\end{tabular} & \begin{tabular}[c]{@{}c@{}}Co-using,\\ Restrictive Mediation\end{tabular}\\
\midrule
\textbf{YP3}  & 10                                         & Girl          & 25-34                                & Female   &  \begin{tabular}[c]{@{}c@{}}Bachelor's\\ Degree\end{tabular}   &  Teacher&\begin{tabular}[c]{@{}c@{}}3-5 Years/\\ Every Day\end{tabular}                             & \begin{tabular}[c]{@{}c@{}}1-3 Years/\\ 1-6 Days Per Week\end{tabular} & \begin{tabular}[c]{@{}c@{}}Restrictive Mediation\end{tabular}\\
\midrule
\textbf{YP4}  & 11                                         & Girl          & 25-34                                & Female   &  \begin{tabular}[c]{@{}c@{}}Master's\\ Degree or Above\end{tabular}    &  Teacher&\begin{tabular}[c]{@{}c@{}}5-7 Years/\\ Every Day\end{tabular}                             & \begin{tabular}[c]{@{}c@{}}3-5 Years/\\ Every Day\end{tabular} & \begin{tabular}[c]{@{}c@{}}Restrictive Mediation\end{tabular}\\
\midrule
\textbf{YP5}  & 12                                         & Boy           & 35-44                                & Female     &  \begin{tabular}[c]{@{}c@{}}Bachelor's\\ Degree\end{tabular}  &  \begin{tabular}[c]{@{}c@{}}Full-time\\ Parent\end{tabular}&\begin{tabular}[c]{@{}c@{}}3-5 Years/\\ 1-6 Days Per Week\end{tabular}                             & \begin{tabular}[c]{@{}c@{}}3-5 Years/\\ Every Day\end{tabular} & \begin{tabular}[c]{@{}c@{}}Active Mediation\end{tabular}\\
\midrule
\textbf{YP6}  & 13                                         & Boy           & 45-54                                & Female    & \begin{tabular}[c]{@{}c@{}}Associate's\\  or Below\end{tabular}   &  \begin{tabular}[c]{@{}c@{}}Full-time\\ Parent\end{tabular}&\begin{tabular}[c]{@{}c@{}}3-5 Years/\\ Every Day\end{tabular}                             & \begin{tabular}[c]{@{}c@{}}1-3 Years/\\ 1-6 Days Per Week\end{tabular} & \begin{tabular}[c]{@{}c@{}}Restrictive Mediation\end{tabular}\\
\midrule
\textbf{YP7}  & 14                                         & Girl          & 35-44                                & Female    &  \begin{tabular}[c]{@{}c@{}}Bachelor's\\ Degree\end{tabular}   &  Teacher&\begin{tabular}[c]{@{}c@{}}3-5 Years/\\ 1-6 Days Per Week\end{tabular}                             & \begin{tabular}[c]{@{}c@{}}1-3 Years/\\ 1-6 Days Per Week\end{tabular} & \begin{tabular}[c]{@{}c@{}}Restrictive Mediation\end{tabular}\\
\midrule
\textbf{YP8}  & 14                                         & Boy           & 35-44                                & Female    &  \begin{tabular}[c]{@{}c@{}}Bachelor's\\ Degree\end{tabular}   &  \begin{tabular}[c]{@{}c@{}}Company\\ Employee\end{tabular}&\begin{tabular}[c]{@{}c@{}}3-5 Years/\\ Every Day\end{tabular}                             & \begin{tabular}[c]{@{}c@{}}1-3 Years/\\ 1-6 Days Per Week\end{tabular} & \begin{tabular}[c]{@{}c@{}}Restrictive Mediation\\Active Mediation\end{tabular}\\
\midrule
\textbf{YP9}  & 14                                         & Girl          & 35-44                                & Female    &  \begin{tabular}[c]{@{}c@{}}Bachelor's\\ Degree\end{tabular}   &  \begin{tabular}[c]{@{}c@{}}Company\\ Employee\end{tabular}&\begin{tabular}[c]{@{}c@{}}5-7 Years/\\ 1-6 Days Per Week\end{tabular}                             & \begin{tabular}[c]{@{}c@{}}1-3 Years/\\ 1-6 Days Per Week Minutes\end{tabular} & \begin{tabular}[c]{@{}c@{}}Restrictive Mediation\\Active Mediation\end{tabular}\\
\midrule
\textbf{YP10} & 14                                         & Boy          & 35-44                                 & Female   &  \begin{tabular}[c]{@{}c@{}}Bachelor's\\ Degree\end{tabular}   &  \begin{tabular}[c]{@{}c@{}}Company\\ Employee\end{tabular}&\begin{tabular}[c]{@{}c@{}}5-7 Years/\\ Every Day\end{tabular}                             & \begin{tabular}[c]{@{}c@{}}1-3 Years/\\ 1-6 Days Per Week\end{tabular} & \begin{tabular}[c]{@{}c@{}}Restrictive Mediation\end{tabular}\\
\midrule
\textbf{YP11} & 15                                         & Girl        & 25-34                                  & Female    & \begin{tabular}[c]{@{}c@{}}Associate's\\ Degree or Below\end{tabular}    &  Nurse&\begin{tabular}[c]{@{}c@{}}3-5 Years/\\ 1-6 Days Per Week\end{tabular}                             & \begin{tabular}[c]{@{}c@{}}3-5 Years/\\ Every Day\end{tabular} & \begin{tabular}[c]{@{}c@{}}Restrictive Mediation\\Active Mediation\end{tabular}\\ \hline
\end{tabular}
}
\end{table*}

\subsubsection{Procedure}
\label{formative_procedure}

We conducted semi-structured interviews via online meetings, separately with each parent and child, lasting 20-30 minutes each. The parent interviews focused on their collaborative approaches with children when using online video platforms, management of existing platform functions and video content, and needs and suggestions for improvement. With parental consent and teenage assent, we interviewed the teenagers about their usage habits and needs for managing video content. All interviews were recorded and transcribed with participant consent, using automated tools. Participants were assured that all interview-related data and materials would remain confidential.

 

We employed the thematic analysis method proposed by Braun and Clarke to code the interview transcripts \cite{braun2006using, brule2020thematic}, following the six-phase process outlined in the codebook thematic analysis \cite{braun2012thematic, braun2013successful, braun2014thematic, braun2021one}. Initially, three authors independently reviewed all participants' feedback multiple times to familiarize themselves with the data and systematically identify key concepts, themes, and patterns for coding. Subsequently, they conducted open coding, producing an initial set of codes that captured relevant themes and patterns emerging from the data. The initial codes were then grouped and categorized into broader themes, reflecting significant patterns relevant to the research objectives. Next, the authors collaboratively reviewed and refined the identified themes, ensuring they were coherent and aligned with the research objectives. Weak or redundant themes were removed, and subthemes were identified when necessary. Moreover, each theme was defined in clear and specific terms, supported by examples from the data, and named to encapsulate its essence. Once the themes were finalized, the authors reviewed the data again to apply the codes comprehensively, selecting illustrative examples to accurately represent each theme and support the findings. Although the six phases are presented in a logical sequence, the thematic analysis is not a strictly linear process. Instead, it is recursive and iterative, requiring the researchers to move back and forth between phases as necessary. Regular communications and collaborative reviews were maintained throughout the process to ensure transparency, accuracy, and reliability. The coding process ended when all authors reached a consensus on the final themes.



\subsection{Results}

\subsubsection{Issues with Current Practices}


Our interviews revealed that parents employed all three types of mediation identified in the PMT framework: restrictive mediation, active mediation, and co-using. Among the parent-child pairs, almost all parents reported using restrictive mediation by controlling video types, particularly for high-risk content such as violence and pornography. Some parents indicated occasional use of active mediation, discussing platform usage with children. Besides, a few parents sometimes adopted co-using by selecting video content before watching together with their children.


However, we found that existing video censorship practices were primarily carried out by parents, heavily relying on their time, knowledge, and experience. Parents reported that monitoring video content for their children was time-consuming and labor-intensive, often requiring them to select appropriate videos while navigating various platform settings tailored for teenagers. As one parent mentioned, ``\textit{I have to watch it first before letting her watch... it's really troublesome}'' (P2). Both parents and teenagers expressed that the filtering of risky content and identification of suitable material did not fully meet their needs. Some risky content could still bypass filters, as noted by one parent: ``\textit{Not all content in the youth mode is suitable}'' (P11). Similarly, a teenager shared, ``\textit{Sometimes I come across gory content}'' (Y3). On the other hand, content identified as suitable often failed to meet teenagers' diverse interests, with one teenager complaining, ``\textit{I can't find what I want to watch}'' (Y6).

\subsubsection{Key Findings} \label{Key Findings}
In response to these issues, we explored parents' and teenagers' expectations regarding video censorship, which can inform more effective approaches to parental mediation. The key findings are summarized below:
\begin{itemize}
    \item \textbf{F1: Video censorship should support filtering out undesirable content and selecting appropriate content in an explainable manner.} Most parents expressed the desire for a video censorship tool to block high-risk content for teenagers. For example, P4 noted that many videos are ``\textit{addictive, risky and low-quality... they should be filtered out}''. Additionally, nearly all parents emphasized the importance of providing suitable videos, with P11 stating, ``\textit{The content should be positive}''. Moreover, both parents and teenagers highlighted the necessity for explainable feedback on detection results. Some parents voiced a desire for additional information to assess content suitability, such as ``\textit{descriptive text}'' (P2) or ``\textit{risk identification and age-appropriate tips}'' (P3). Similarly, some teenagers indicated a desire for prompts to help evaluate videos, as one said, ``\textit{It would help me decide if it's appropriate}'' (Y7).


    \item \textbf{F2: Parents and teenagers expect video censorship to balance common requirements and personal preferences.} Parents unanimously agreed on the need to block content involving violence, pornography, and negative themes, but their specific concerns varied. For example, P9 was worried about ``\textit{gory violence}'', while P10 was more concerned with ``\textit{entertainment content}''. Similarly, while all parents preferred educational content, the specific types differed. For instance, P11 mentioned content ``\textit{related to history and politics}'', whereas P2 preferred content ``\textit{focused on art and news}''. Teenagers also expressed their own preferences, with Y4 stating, ``\textit{I like astronomy and geography videos}''.

    \item \textbf{F3: The personalized censorship needs are generally generated collaboratively by parents and teenagers through mediation.} Most parents wanted to be involved in configuring video content for their children, expressing wishes to ``\textit{set the types of videos}'' (P3) and ``\textit{limit him from videos related to games}'' (P8). They also acknowledged the necessity of allowing teenagers to express their preferences, as P5 explained, ``\textit{It's necessary to allow him to do some settings himself}''. Although some teenagers indicated openness to parental involvement, they generally preferred to avoid direct discussions, fearing misunderstandings or conflicts. Y9 mentioned, ``\textit{I don't dare to tell my parents face-to-face, because they don't always understand my interests}''. They also valued their privacy and independence, as Y5 said, ``\textit{It's my personal thing}''. Moreover, teenagers felt that direct negotiations were often unproductive, feeling that discussions ``\textit{usually end up going their ways}'' (Y10). Consequently, teenagers favored indirect collaboration, such as sharing videos via media platforms, as Y3 mentioned, ``\textit{I send my parents videos I find interesting so they can see what I've been into lately}''.


    \item \textbf{F4: Besides directly expressing censorship needs, parents' and teenagers' preferences are usually hidden in some videos.} Parents and teenagers sometimes clearly expressed their video preferences, such as ``\textit{directly choose videos in the anime section}'' (Y11). However, they found it difficult in some circumstances. In such cases, censorship preferences were often conveyed through videos they had collected or shared. Some parents mentioned that they often selected suitable content for their children by collecting and sharing videos. For example, P10 stated, ``\textit{Sometimes I saved a video and showed it to my child when it's convenient}''. Similarly, video selection is a primary way for children to express their interests and desires to their parents. As one parent noted, ``\textit{The videos she sends help me understand what she is interested in}'' (P9). 


    \item \textbf{F5: Parents expect the summarization of teenagers' censorship results.} Most parents expressed curiosity about their children's activity trajectories. For example, P1 mentioned, ``\textit{I sometimes check her viewing history}''. However, a few parents recognized that this might harm the parent-child relationship, as it could feel intrusive to children, with P5 explained, ``\textit{He deletes his history after using}''. To address this, parents suggested to ``\textit{provide summaries, such as video types have been watched}'' (P2). Some parents expected such summaries to be available when needed, as P5 explained, ``\textit{It would be useful if I could access it when I want to know}''.
\end{itemize}
  
\subsubsection{Design Goals}
\label{Design Goals}
Based on the findings, we propose five design goals for a parent-child video censorship tool:

\begin{itemize}
    \item \textbf{D1: Conduct video censorship from both risk and suitability perspectives (F1).} The automated video censorship tool should address parents' and teenagers' needs for risk filtering and appropriateness identification.

     \item \textbf{D2: Support collaborative configuration of personalized censorship needs through mediation (F2, F3).} Mediation is suggested to be introduced to enable collaboration and interaction between parents and teenagers when configuring their personalized censorship needs.

    \item \textbf{D3: Support both direct and indirect methods for configuring personalized censorship needs (F3, F4).} The personalized video censorship process can be set up through direct manual configuration (e.g., selecting video content types and their weights) and indirect automatic configuration (e.g., learning preferred content features based on the videos selected by users).

    \item \textbf{D4: Provide both in-time and long-term explanatory feedback on censorship needs (F1, F5).} The censorship tool should offer immediate explanations regarding personalized censorship needs to guide video selection for teenagers and parents, as well as generate long-term summary reports to help parents understand teenagers' usage patterns.

    \item \textbf{D5: Ensure user control (F2, F4, F5).} The tool should be designed to enhance the efficiency and effectiveness of the parent-child JME process in video content censorship, ensuring that both parents and teenagers can participate in and control it.
\end{itemize}

\section{YouthCare: A Personalized Collaborative Video Censorship Tool }
\label{YouthCare}
Based on the design goals, we developed YouthCare, an automated video censorship tool for parents and children to collaboratively filter and select video content in JME. By utilizing the MLLMs, YouthCare enables a joint understanding of user configuration and video content. Next, we will detail how YouthCare is designed and implemented.   

\subsection{YouthCare Design}
\begin{figure}[htbp]
  \centering
  \includegraphics[width=0.70\linewidth]{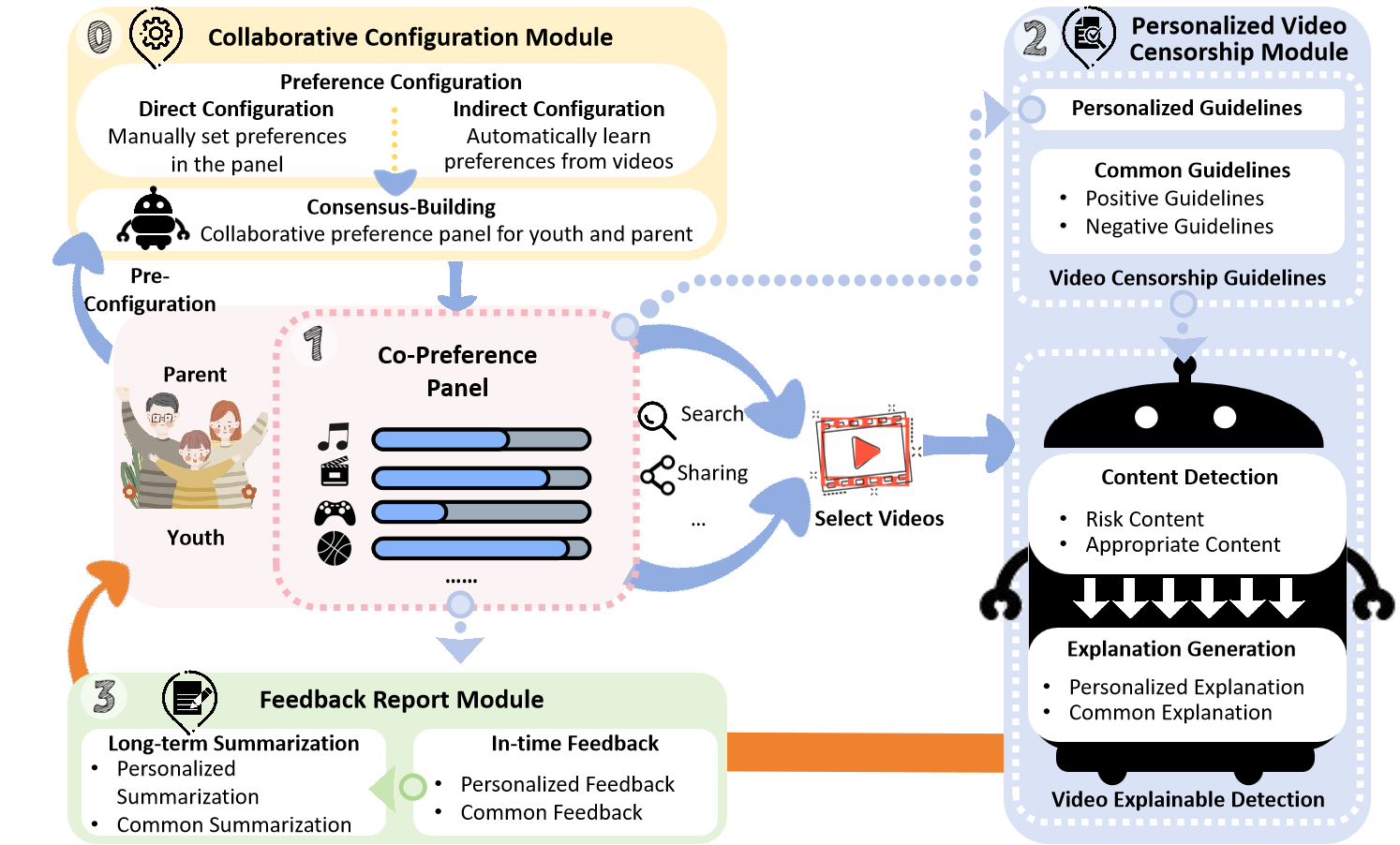}
  \caption{The framework of YouthCare.}
  \label{fig:overview}
\end{figure}

YouthCare consists of three main modules, as shown in Figure~\ref{fig:overview}:

\begin{itemize}
    \item \textbf{Collaborative Configuration Module}: This module allows parents and teenagers to collaboratively configure preferences through both direct and indirect methods. It introduces a Chatbot-based component for consensus-building to reduce conflicts between children and parents. This module generates a ``co-preference'' from the child and parent, which serves as personalized guidance for video censorship and feedback reports.
    
    \item \textbf{Personalized Video Censorship Module}: This module utilizes MLLMs to provide explainable video censorship according to video censorship guidelines. It is achieved by incorporating Personalized Guidelines - based on co-preference, and Common Guidelines - derived from authoritative standards and policies. As such, this module offers both a personalized and standardized basis for content censorship.
    
    \item \textbf{Feedback Report Module}: This module provides in-time feedback on video censorship results based on co-preference. Additionally, it regularly summarizes these in-time censorship results to generate personalized reports for long-term use.
\end{itemize}
\subsubsection{Collaborative Configuration Module}
\label{Collaborative Configuration Module}
Our formative study revealed that parents and teenagers expect to participate collaboratively in personalized video censorship by both direct and indirect methods (F2, F3, F4). Then, we designed the Collaborative Configuration Module to facilitate the collaborative configuration of personalized video censorship needs (D2, D3, D5). This module supports direct manual configuration, such as customizing video keywords and assigning weights, as well as indirect automatic configuration, which infers preferred features from users' video selections. It then assists parents and teenagers in establishing a ``co-preference'' through a Chatbot interface.

\begin{figure}[htbp]
  \centering
  \includegraphics[width=0.8\linewidth]{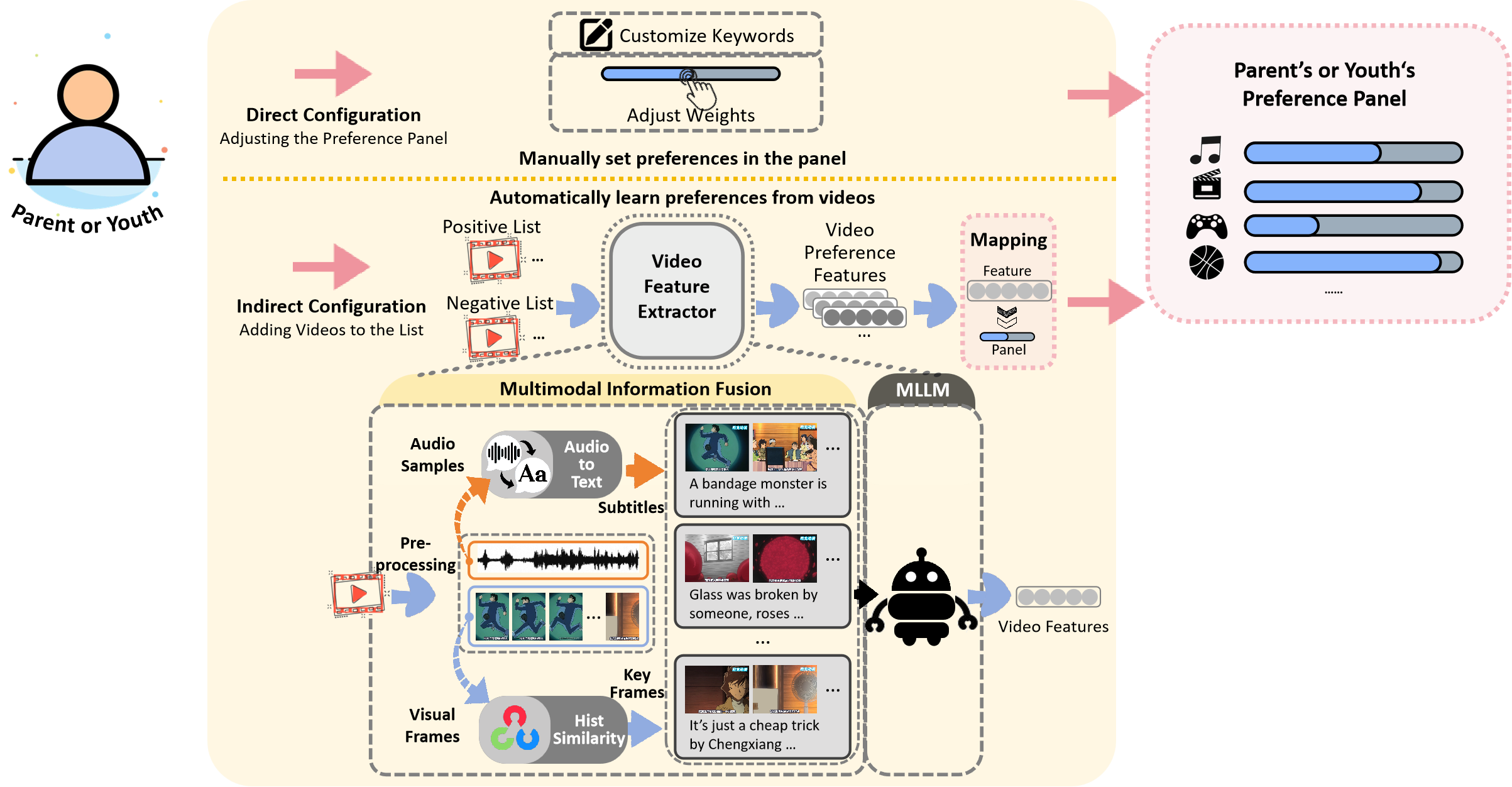}
  \caption{Preference configuration process.}
  \label{fig:model1}
\end{figure}

The module involves two key processes: preference configuration and consensus-building. First, during the preference configuration, parents and teenagers interact with YouthCare through direct and indirect methods to generate personalized content preferences, as shown in Figure~\ref{fig:model1}.

\begin{itemize}

    \item \textbf{Direct Configuration:} This method allows users to directly set keywords and adjust corresponding weights in the preference panel, enabling flexible filtering or selection of video content. Parents or teenagers can customize keywords (e.g., education, violence) and assign weights according to their preferences. The weights are set on a scale of [-2, -1, 0, 1, 2], representing ``strongly dislike'', ``dislike'', ``neutral'', ``like'', and ``strongly like'', respectively. This enables precise preference expression for video content. For example, a child can set a weight of -2 for the keyword ``violence'' to avoid such content or a weight of 2 for ``science'' to prioritize educational videos.

    \item \textbf{Indirect Configuration:} This method implicitly learns user preferences by analyzing the videos that users select. Parents or teenagers can label videos as suitable (preferred content or content they want to watch) or unsuitable (risky content or content they do not want to watch), adding them to positive or negative video list. The Video Feature Extractor then analyzes the content to extract video features. Given the multimodal nature of video content, we employ a Multimodal Information Fusion process combined with an MLLM to parse video content. The video is preprocessed by decomposing it into audio samples and visual frames. Audio samples are converted into timestamped text using an audio-to-text tool, serving as subtitles. We compute frame similarity and extract key frames from visual frames when similarity falls below the threshold. We then align the subtitles with key frames according to timestamps and feed them into the MLLM, along with a carefully designed prompt (Appendix \ref{Appendix_Prompts}), to obtain corresponding video features. Finally, users' preferences are ultimately reflected in the parent's or youth's preference panel, presented as a visual interface highlighting video keywords and weights.

\end{itemize}


The configuration process offers users the flexibility to express their preferences using both direct and indirect methods. Users can choose either method or combine them. For example, they can build initial preferences through direct configuration and then refine them using the indirect method, or first automatically generate preferences through the indirect method and then make manual adjustments based on the results.

\begin{figure}[htbp]
  \centering
  \includegraphics[width=0.9\linewidth]{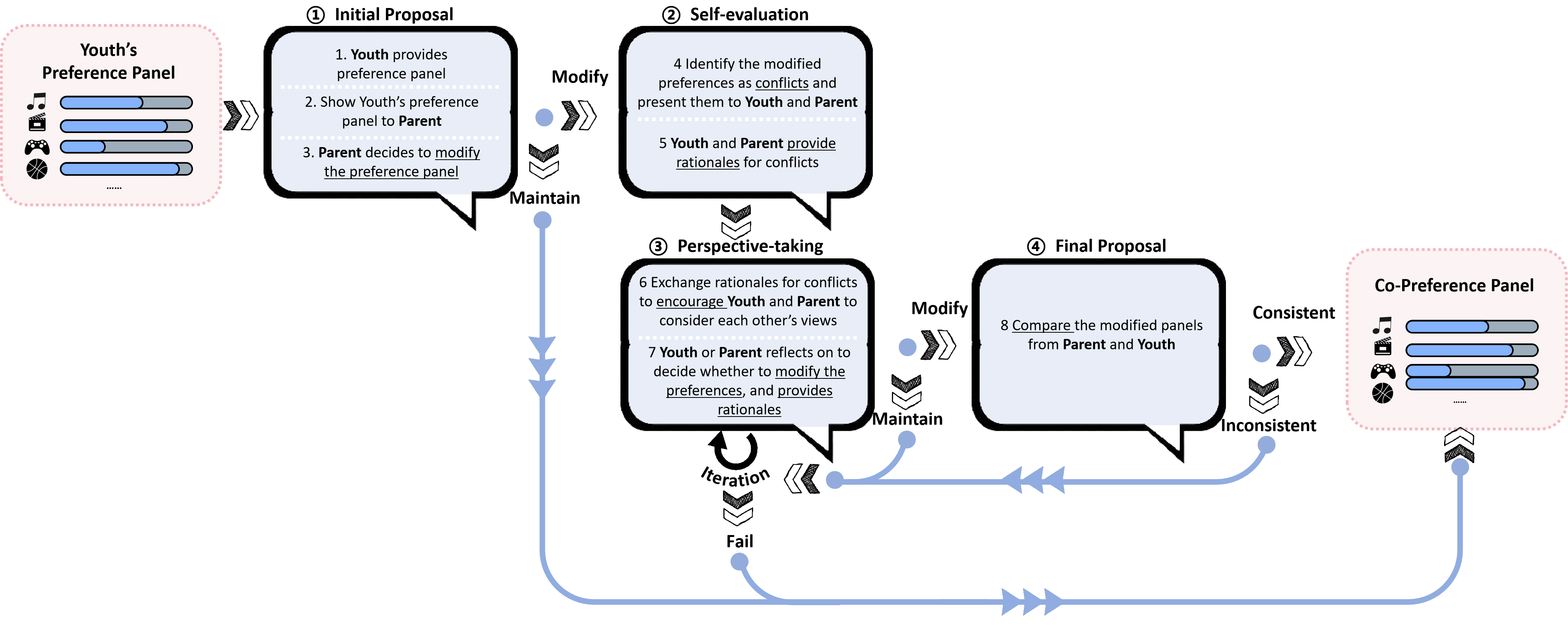}
  \caption{Consensus-building process (illustrating an example where a teenager first provides their preference panel).}
  \label{fig:collaborative}
\end{figure}

Second, we move into the consensus-building process, where the preferences configured by either the teenager or the parent are collaboratively reviewed and adjusted by the other party. As noted in F3 (Section~\ref{Key Findings}), teenagers often prefer to collaborate with parents indirectly through mediation. The goal of this mediation is to achieve consensus-building, defined as ``an attempt at converging views toward a collectively agreed decision'' \cite{andersen2021conflicts,beattie2017generating,son2012consumer}. To enhance the efficiency of this process, we draw on the approach from \cite{shin2022chatbots}, utilizing a Chatbot as a mediator to facilitate consensus-building through conversation. Specifically, the Chatbot guides parents and teenagers through a four-stage dialogue to reach an agreement on preference configuration, with the results displayed in the ``Co-Preference Panel''. We illustrate an example where a teenager first provides the preference panel in Figure ~\ref{fig:collaborative}.

\begin{itemize}
    \item \textbf{Initial Proposal:} Either a parent or a teenager initiates the configuration process by providing their preference panel as described in the above preference configuration process. The Chatbot then presents this preference panel to the other party and guides them in deciding if adjustments are needed. For example, if a teenager first provides his/her preference panel, the Chatbot will forward it to the parent for possible modifications, such as adjusting keywords or weights. Similarly, the parent can also show their configured panel to the child, who can express desired adjustments to parts that do not align with their preferences. If no changes are needed, the preference panel is considered aligned, indicating a successful consensus-building process, and it becomes the final ``Co-Preference Panel''. If modifications are required, the process moves to the next stage for further collaboration.

    \item \textbf{Self-evaluation:} This process begins when the Chatbot identifies and displays conflicts based on the modifications suggested in the previous stage. For example, if the parent or teenager wants to modify keywords, this indicates conflicts regarding those keywords. The Chatbot collects all identified conflicts and presents them separately to both the parent and teenager, and guides them to explain the reasons behind their settings.

    \item \textbf{Perspective-taking:} In this stage, the Chatbot presents each party's reasons for conflicts to the other and guides them to consider each other's viewpoints. Both the parent and teenager then reflect on these conflicts and decide whether to adjust the settings, providing reasons for any changes. If either party compromises, the process proceeds to the next stage; otherwise, the Chatbot re-presents the explanations and repeats this stage until a change is agreed upon. If the maximum number of iterations is reached without an agreement, it indicates a failure to reach the consensus, and the existing configuration panel becomes the final ``Co-Preference Panel''.
    

    \item \textbf{Final Proposal:} The Chatbot compares the updated preference panels from both the parent and teenager to check for consensus-building. If there are still inconsistencies, it will return to the previous Perspective-taking stage and re-present the explanations for further discussion. Otherwise, the panel is established as the final ``Co-Preference Panel'', and the consensus-building process is finished.

\end{itemize}

Through these four steps, the final ``Co-Preference Panel'', collaboratively configured by the parent and teenager via the Chatbot, is built. This panel provides personalized support for the Personalized Guidelines in the Personalized Video Censorship Module, and for the feedback and summarization generation in the Feedback and Report Module, which will be detailed in subsequent sections.

\subsubsection{Personalized Video Censorship Module}
\label{Personalized Video Censorship Module}
\begin{figure}[htbp]
  \centering
  \includegraphics[width=0.90\linewidth]{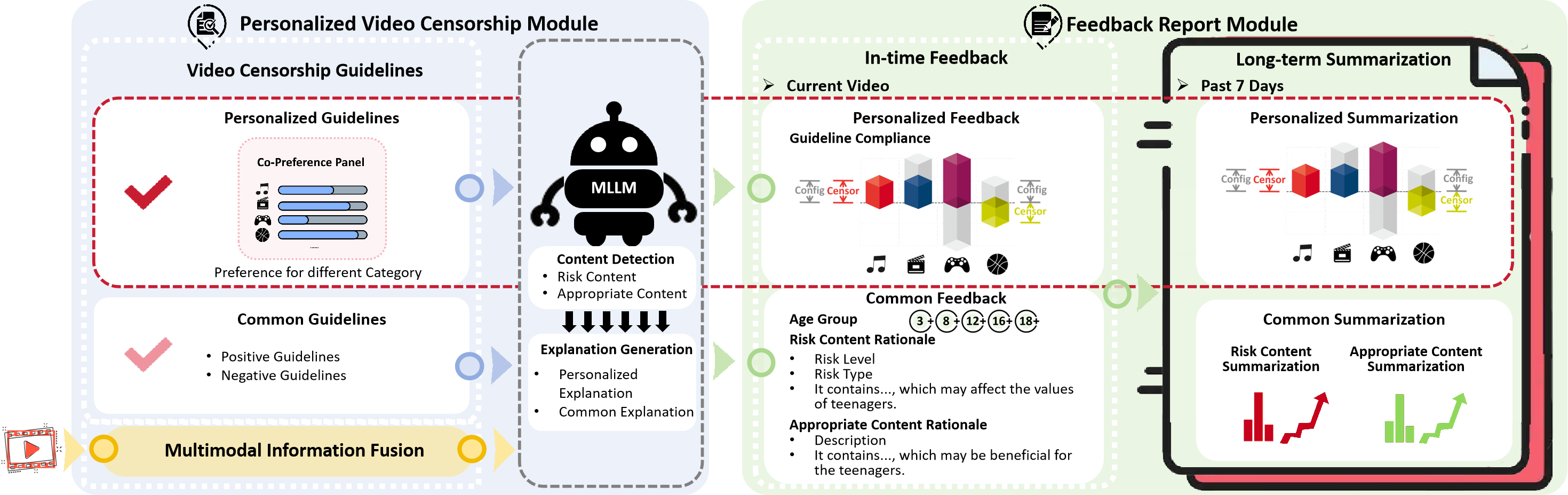}
  \caption{Personalized Video Censorship Module and Feedback Report Module.}
  \label{fig:model2}
\end{figure}

The formative study indicates that parents and teenagers prefer a video censorship system that can filter and select content, while providing personalized and general explanations (F1, F2). Therefore, we designed the Personalized Video Censorship Module based on MLLMs to achieve explainable censorship in YouthCare (D1). As shown in the left part of Figure~\ref{fig:model2}, this module integrates Multimodal Information Fusion and an MLLM, guided by Video Censorship Guidelines to support both personalized and standardized censorship. 

Specifically, the Multimodal Information Fusion process is similar to the indirect configuration described in Section~\ref{Collaborative Configuration Module} (Figure~\ref{fig:model1}), resulting in the extraction of video features. MLLM subsequently analyzes these features to assess the video's risk and appropriateness, providing corresponding explanations. To generate effective explanations, we use a guideline-based approach to help the model better understand the censorship results \cite{liang2024guide}. By defining two types of Video Censorship Guidelines - Personalized Guidelines and Common Guidelines, we aim to ensure that the video censorship results are both personalized and standardized.

\begin{itemize}
    \item \textbf{Personalized Guidelines:} These guidelines provide personalized explanations for the detection by matching the extracted video features, such as the video's keywords and weights, with the user's preferences. The guidelines are derived from the ``Co-Preference Panel'', which is established through parents and teenagers in the Collaborative Configuration Module. This ensures that the detection results accurately reflect the specific needs of users.
    
    
    \item \textbf{Common Guidelines:} These guidelines offer standardized criteria for video censorship, focusing on age appropriateness, risk, suitability, and other factors. The guidelines are based on authoritative standards and policies, providing comprehensive norms. Specifically, common guidelines address two aspects: video content and youth considerations. For video content, they include standards for film and television, such as the Motion Picture Association film rating system \cite{mpaa2007} and Federal TV rating guidelines \cite{tv_rating}. For youth considerations, the guidelines include various domestic and international standards regarding teenagers' exposure to online content. Examples include \textit{the Guidelines for the Construction of Mobile Internet Minor Mode (Draft for Comments)} \cite{cac2023} and the \textit{Classification and Codes of Unhealthy Internet Content for Adolescents} \cite{cy2019}. In practice, these guidelines can be dynamically adjusted for various scenarios. 
\end{itemize}



\subsubsection{Feedback Report Module}
\label{Feedback Report Module}
According to the formative study, both parents and teenagers are concerned with censorship explanations, while parents particularly focus on summary regarding teenagers' usage (F1, F5). Therefore, we designed the Feedback Report Module to provide both in-time and long-term explanatory feedback based on users' censorship needs (D4). As shown in the right part of Figure~\ref{fig:model2}, this module offers a comprehensive understanding of video censorship, including in-time feedback for current video and long-term summary over a specific period.

\begin{itemize}
    \item \textbf{In-time Feedback:} After a selected video finishes detection, the system displays two types of feedback in real-time. First, based on the Personalized Guidelines, the system calculates how well the features of the current video align with the users' ``co-preference''. This alignment is scored on a scale of [-2, -1, 0, 1, 2], representing ``very low'', ``low'', ``medium'', ``high'', and ``very high'', respectively, for how much the video matches specific keywords. 
   For example, if the preference for the keyword ``Music'' on the ``Co-Preference Panel'' is set to 1 (representing ``like''), and the current video scores 1 (representing ``high'') for the keyword ``Music'', it indicates that the video aligns well with the user's preference for this type. This result is visually presented in a bar chart format. Second, the system generates detection results based on the Common Guidelines, which include the appropriate age group, potential risk content (e.g., risk level and type), appropriate content (e.g., educational or entertainment value), and detailed descriptions of such results.

    \item \textbf{Long-term Summarization:} Over the detection period, data from each in-time feedback are aggregated into a comprehensive statistical report. The report also includes two aspects. First, for the personalized summarization, the report provides the average values for keywords detected during this period and their alignment with the ``Co-Preference Panel''. For example, if 10 videos are detected over a week and the average rating for the keyword ``Games'' is 2 (representing ``very high''), while the co-preference for it is set to -2 (representing ``strongly dislike''), this reveals a mismatch between user preferences and the content. Second, from the perspective of common summarization, the report visually analyzes the common features of videos regarding risk and appropriateness during the period. For example, it uses bar charts to show the frequency distribution of different risk types and line graphs to depict the trends in the frequency of each risk type over time.

\end{itemize}

The Feedback Report Module offers two key advantages. First, in-time feedback reduces the burden on parents in selecting appropriate content and assists teenagers in choosing content that better suits their preferences. Second, long-term summarization provides parents with a comprehensive understanding of the quality of content their child has viewed, allowing them to offer targeted guidance while respecting the child's privacy.

\subsubsection{Example Process}
\label{Example Process}
We detail the three phases of using YouthCare, as shown in Figure~\ref{fig:overview}.

\begin{itemize}
    \item \textbf{Pre-Configuration Process:} Before using YouthCare for video censorship, parents and teenagers collaboratively pre-configure their video preferences through the Collaborative Configuration Module. Configuration can be initiated directly (e.g., customizing video keywords and weights) or indirectly (e.g., selecting videos) by the parent or the teenager. A Chatbot then facilitates consensus-building to establish a ``co-preference''. This setup, applied initially or when preferences change, is automatically used for all subsequent detection, reducing repetitive setup and enhancing usability. 

    \item \textbf{Immediate Usage Process:} While using YouthCare, users select videos by searching or sharing. The Personalized Video Censorship Module then analyzes the content using MLLMs to identify risky and appropriate content. Before deciding to watch, users can view in-time feedback from the Feedback Report Module, which provides both personalized and general feedback.

    \item \textbf{Long-term Review Process:} YouthCare also supports long-term usage by providing parents with detailed summaries of videos watched by teenagers over periods. These summaries, available at any time, include both personalized and common censorship analysis, with visual statistics showing alignment with the ``co-preference''. This enables parents to monitor YouthCare's effectiveness in real-time, understand children's media usage patterns, and evaluate content alignment with preferences and safety guidelines while protecting children’s privacy by focusing on overall trends rather than specific video details.
\end{itemize}

\subsection{YouthCare Implementation}
We selected WeChat for implementing YouthCare due to its wide usage in China, particularly among teenagers and parents, and its support for intuitive text-based communication within families \cite{luo2019using}. The YouthCare Agent connects to WeChat via the WeChaty module, functioning as a Chatbot account linked to a backend program. 
Upon activation, parents and teenagers add the Agent as a friend and enter a pre-generated verification code to complete the parent-child pairing. 
Until both parties have added the Agent, the Agent automatically sends a welcome message to provide a brief introduction to help them quickly understand and start using the system (Figure~\ref{fig:get initial panel} (a)).


\subsubsection{Preference Configuration Process}

\begin{figure*}[ht]
    \begin{minipage}[t]{0.33\textwidth}
        \centering
        \includegraphics[width=\textwidth]{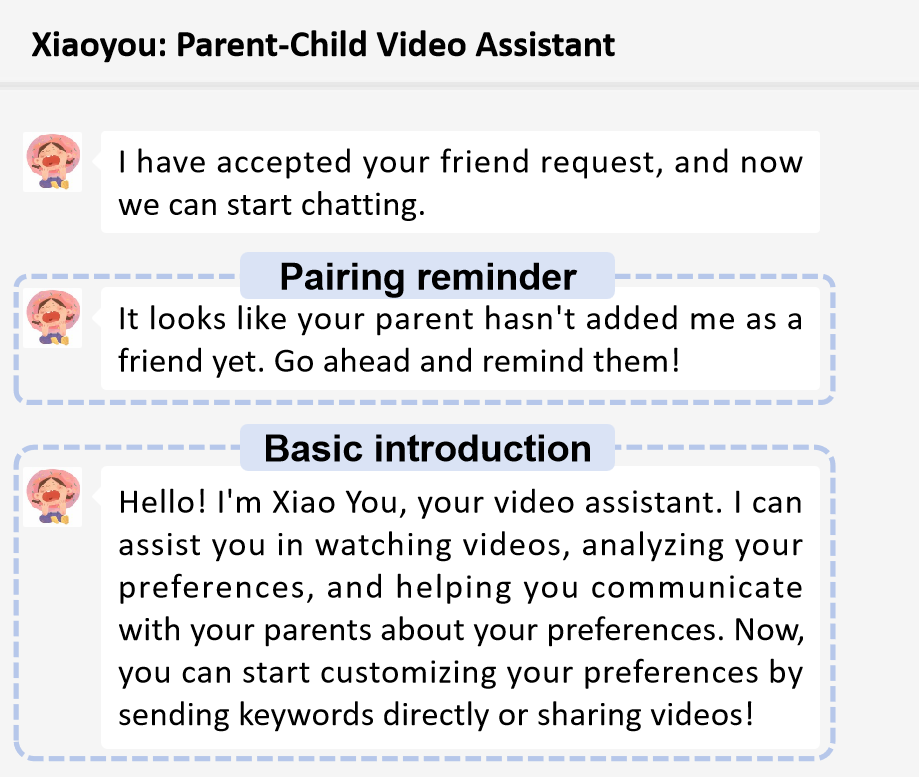}
        \label{fig:initiao}
        \centerline{\small{(a) Initiate YouthCare agent.}}
    \end{minipage}%
    \hfill
    \begin{minipage}[t]{0.33\textwidth}
        \centering
        \includegraphics[width=\textwidth]{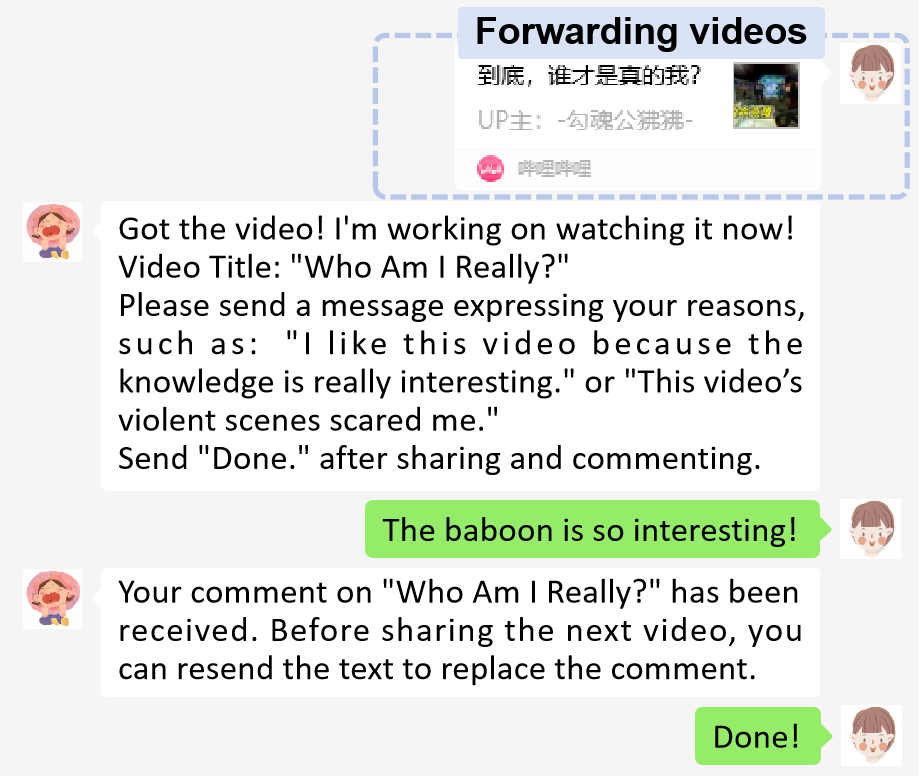}
        \label{fig:6-a}
        \centerline{\small{(b) Indirect configuration.}}
    \end{minipage}%
    \hfill
    \begin{minipage}[t]{0.33\textwidth}
        \centering
        \includegraphics[width=\textwidth]{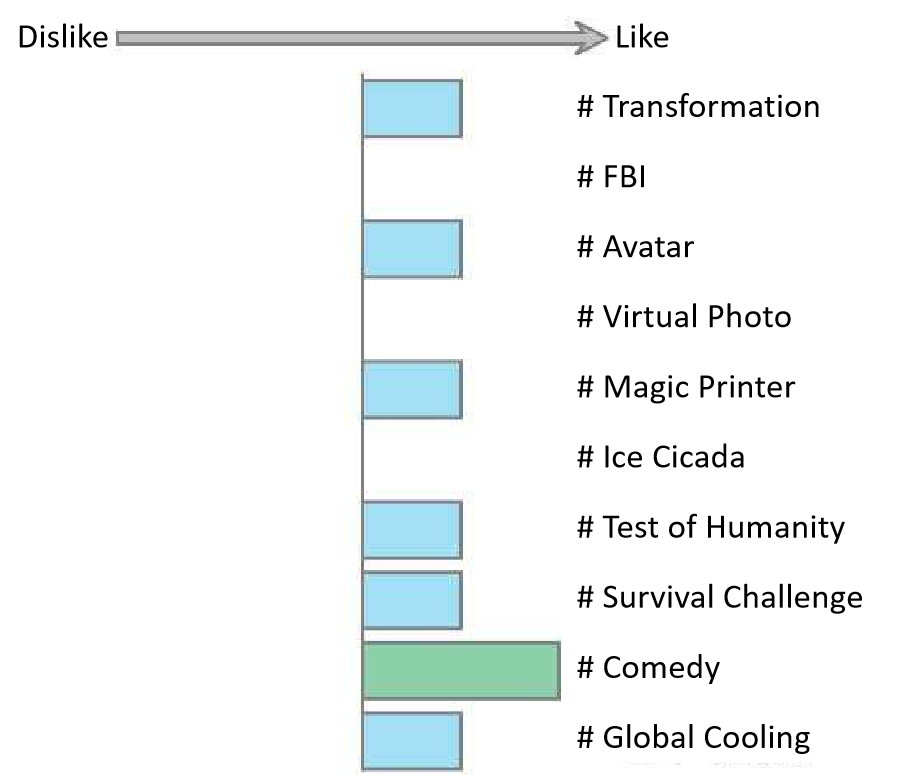}
        \label{fig:6-b}
        \centerline{\small{(c) Initial preference panel.}}
    \end{minipage}
    \vfill
    \caption{Get initial preference panel (the interface is translated from Chinese for clarity).}
    \label{fig:get initial panel}
\end{figure*}

\begin{figure*}[t]
    \begin{minipage}[t]{0.33\linewidth}
        \centering
        \includegraphics[width=\textwidth]{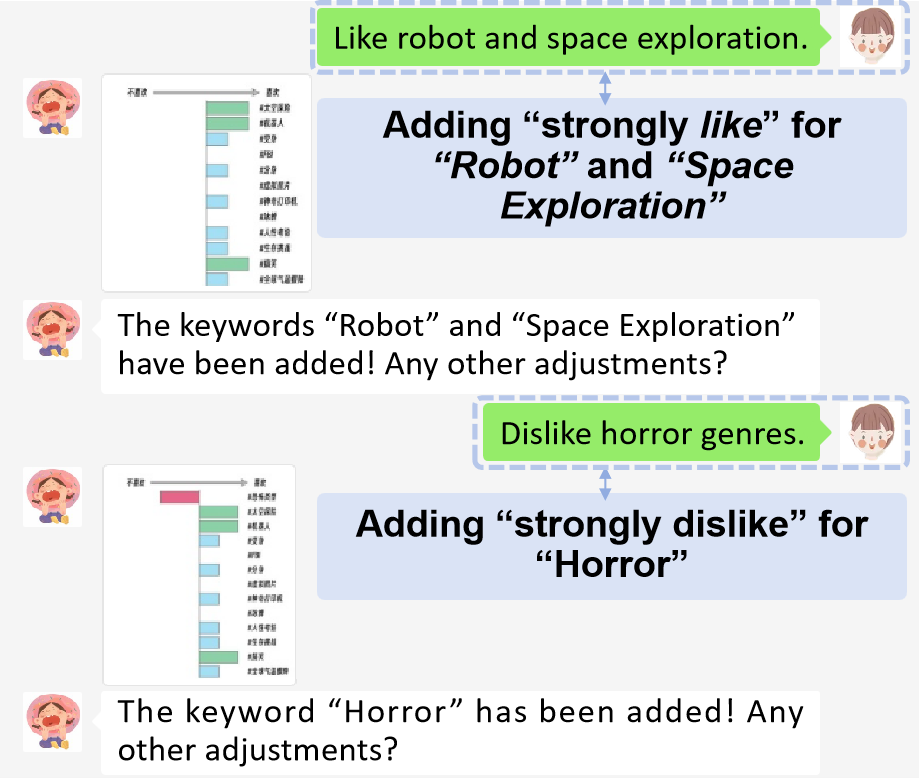}
        \centerline{\small{(a) Direct configuration.}}
    \end{minipage}%
    \hfill
    \begin{minipage}[t]{0.33\linewidth}
        \centering
        \includegraphics[width=\textwidth]{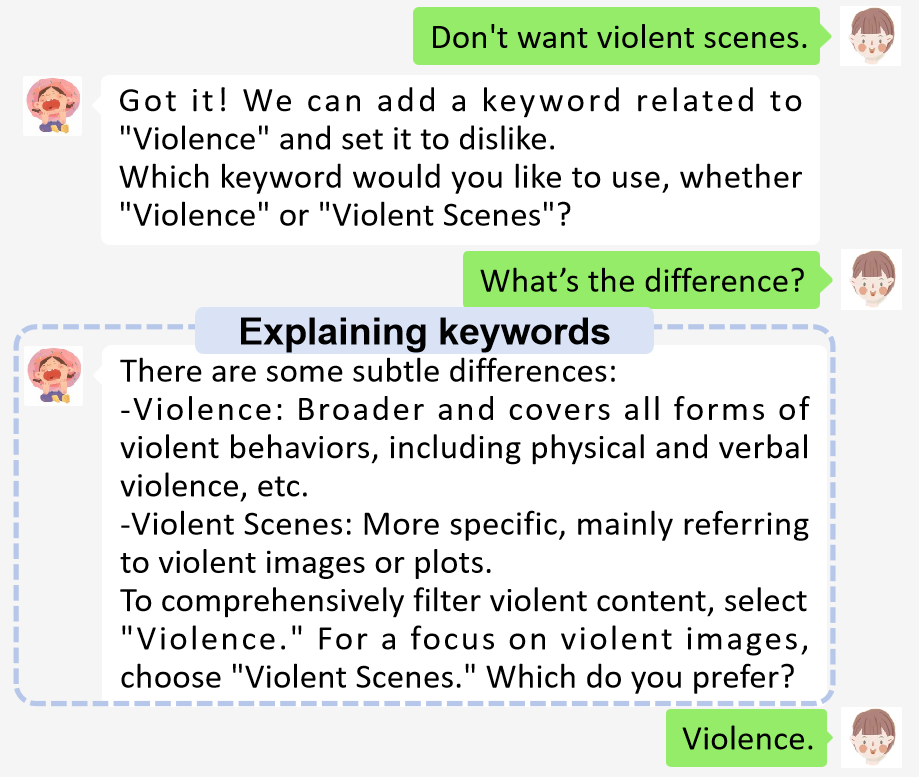}
        \centerline{\small{(b) Assist in understanding keywords.}}
    \end{minipage}
    \hfill
    \begin{minipage}[t]{0.33\linewidth}
        \centering
        \includegraphics[width=\textwidth]{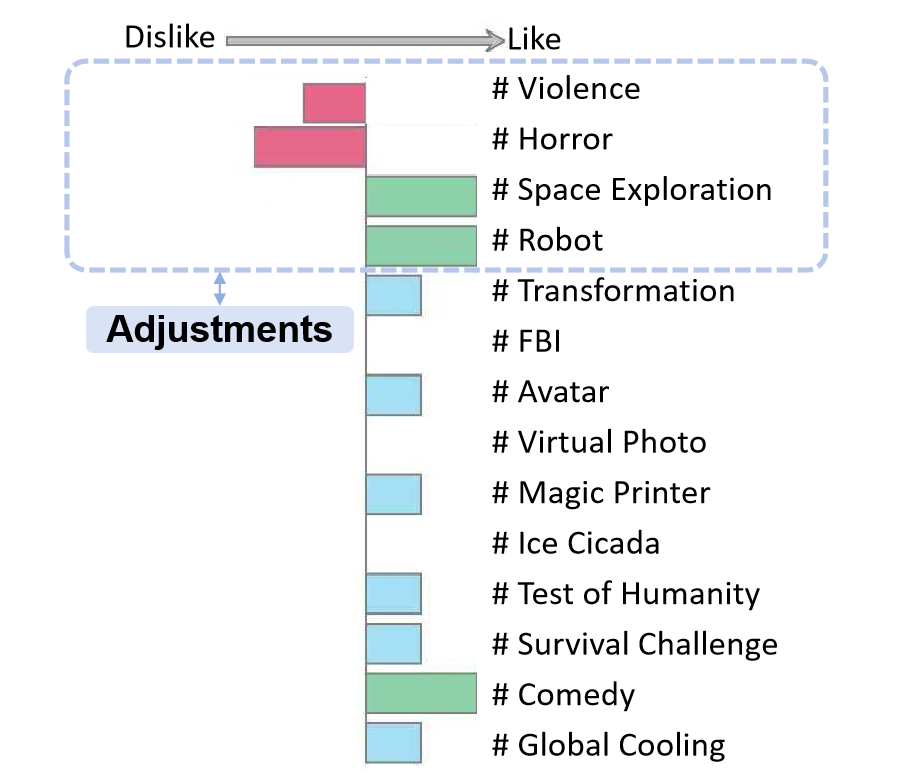}
        \centerline{\small{(c) Final preference panel for teenager.}}
    \end{minipage}
    \vfill
    \caption{Modify preference panel by the teenager (the interface is translated from Chinese for clarity).}
    \label{fig:Modify preference panel}

\end{figure*}

We take an example of a configuration initiated by teenagers, but the same process applies to parents. As illustrated in Figure~\ref{fig:get initial panel} (b), the teenager begins by forwarding videos from existing video platforms on WeChat and providing preference for each. After forwarding is completed, the Agent employs iFLYTEK's ASR service\footnote{https://global.xfyun.cn/products/real-time-asr} and the GPT-4o model\footnote{https://openai.com/index/hello-gpt-4o/} to sequentially analyze the videos. Then, the system presents the teenager's preference keywords and corresponding weights on a visual panel (Figure~\ref{fig:get initial panel} (c)). 
Afterward, the teenager can directly modify the current preference panel according to conversations as Figure~\ref{fig:Modify preference panel} (a) and Figure~\ref{fig:Modify preference panel} (b), with the system dynamically updating the panel as Figure~\ref{fig:Modify preference panel} (c), until no further adjustments are needed.

\subsubsection{Collaborative Configuration Process}

\begin{figure*}[t]
    \begin{minipage}[t]{0.36\linewidth}
        \centering
        \includegraphics[width=\textwidth]{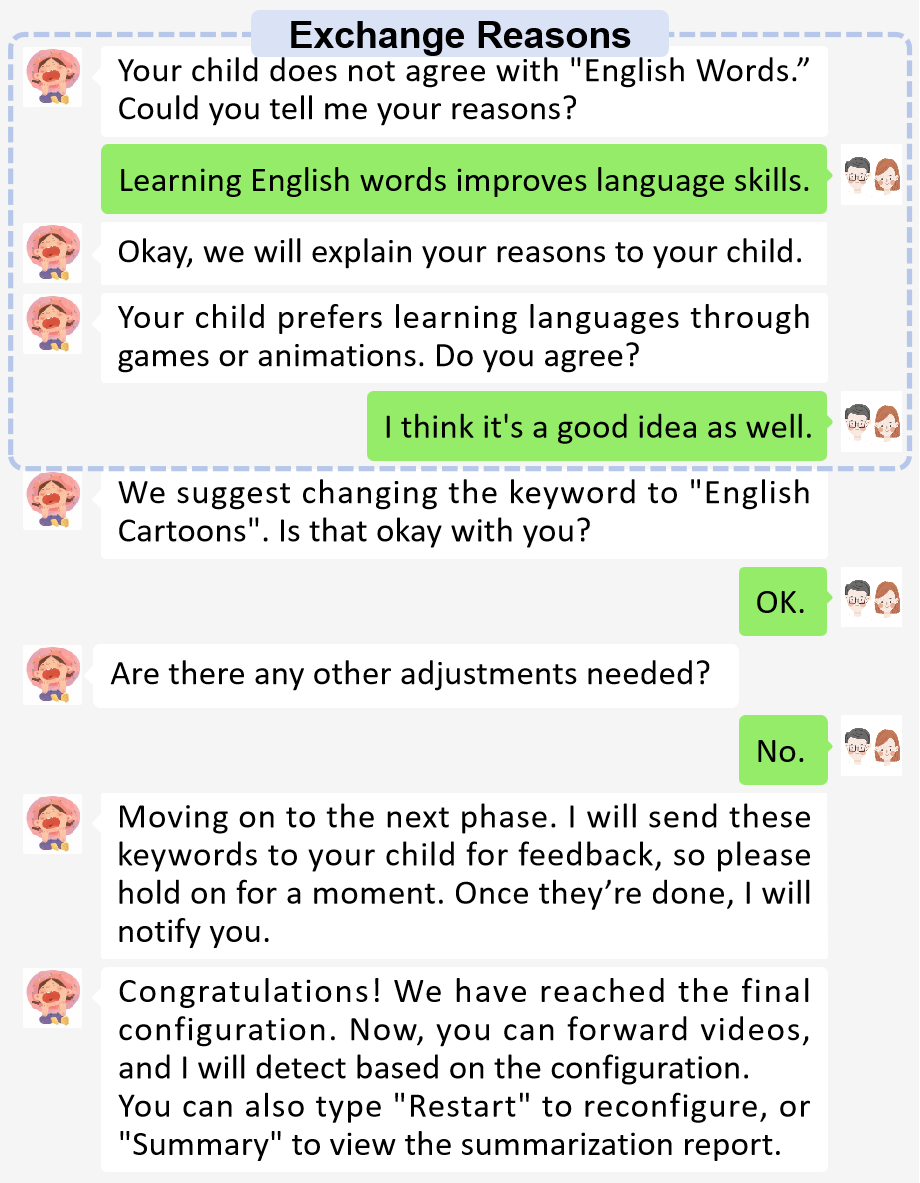}
        \centerline{\small{(a) Assist parent in understanding teenager's view.}}
    \end{minipage}
    \hspace{2cm}
    \begin{minipage}[t]{0.36\linewidth}
        \centering
        \includegraphics[width=\textwidth]{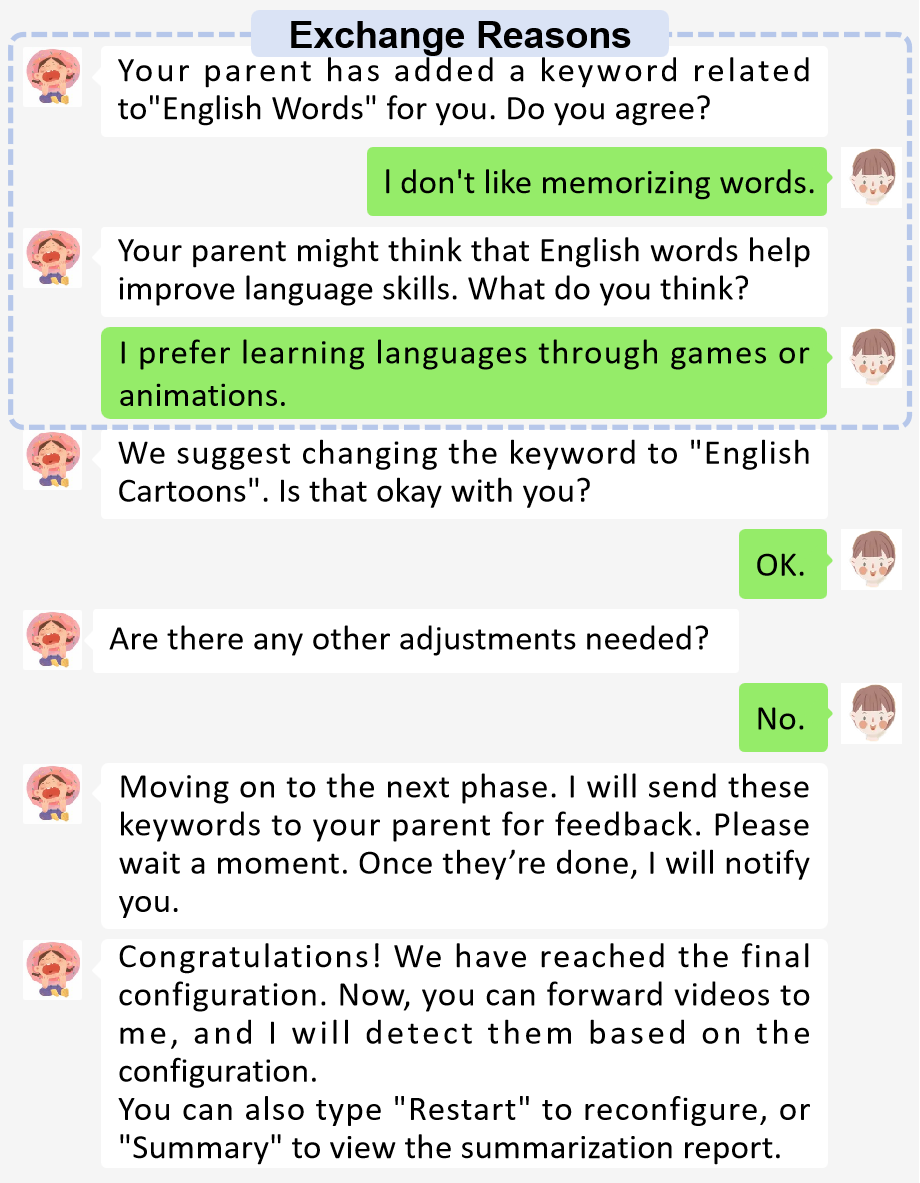}
        \centerline{\small{(b) Assist teenager in understanding parent's view.}}
    \end{minipage}
    \caption{Assist in understanding each other's views (the interface is translated from Chinese for clarity).}
    \label{fig:Understand each other's view}

\end{figure*}

Then, parents and teenagers enter the collaboration process. When the parent receives the child's panel, he/she can also use dialogue with the Agent to express opinions. 
The Agent acts as a neutral mediator, which facilitates communication between both parties and encourages them to explain their reasons (Figure~\ref{fig:Understand each other's view} (a) and Figure~\ref{fig:Understand each other's view} (b)). 
The Agent conveys each party's reasons to foster reflection and understanding. This process continues until both parties agree and have no further modification requests, and then, the system generates a Co-Preference Panel that reflects their mutual content preferences.

\subsubsection{Feedback Process}
\begin{figure*}[t]
    \begin{minipage}[t]{0.37\linewidth}
        \centering
        \includegraphics[width=\textwidth]{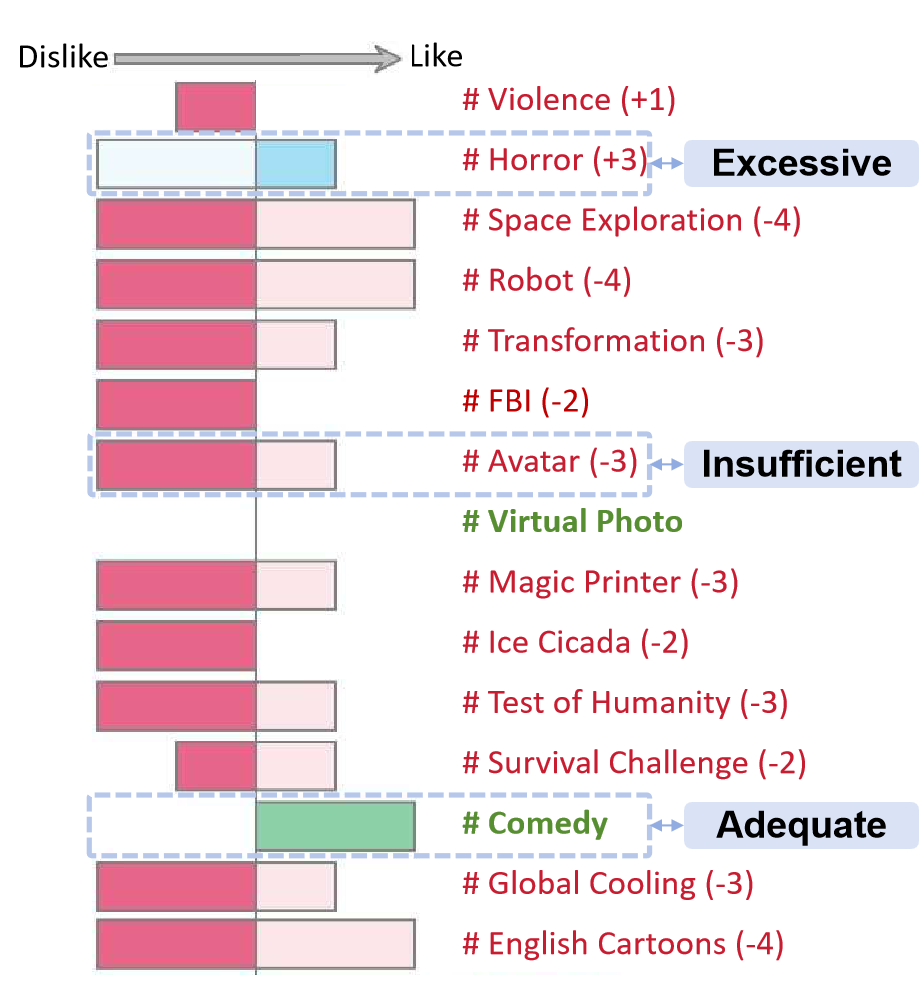}
        \centerline{\small{(a) Personalized in-time feedback.}}
    \end{minipage}%
    \hfill
    \begin{minipage}[t]{0.37\linewidth}
        \centering
        \includegraphics[width=\textwidth]{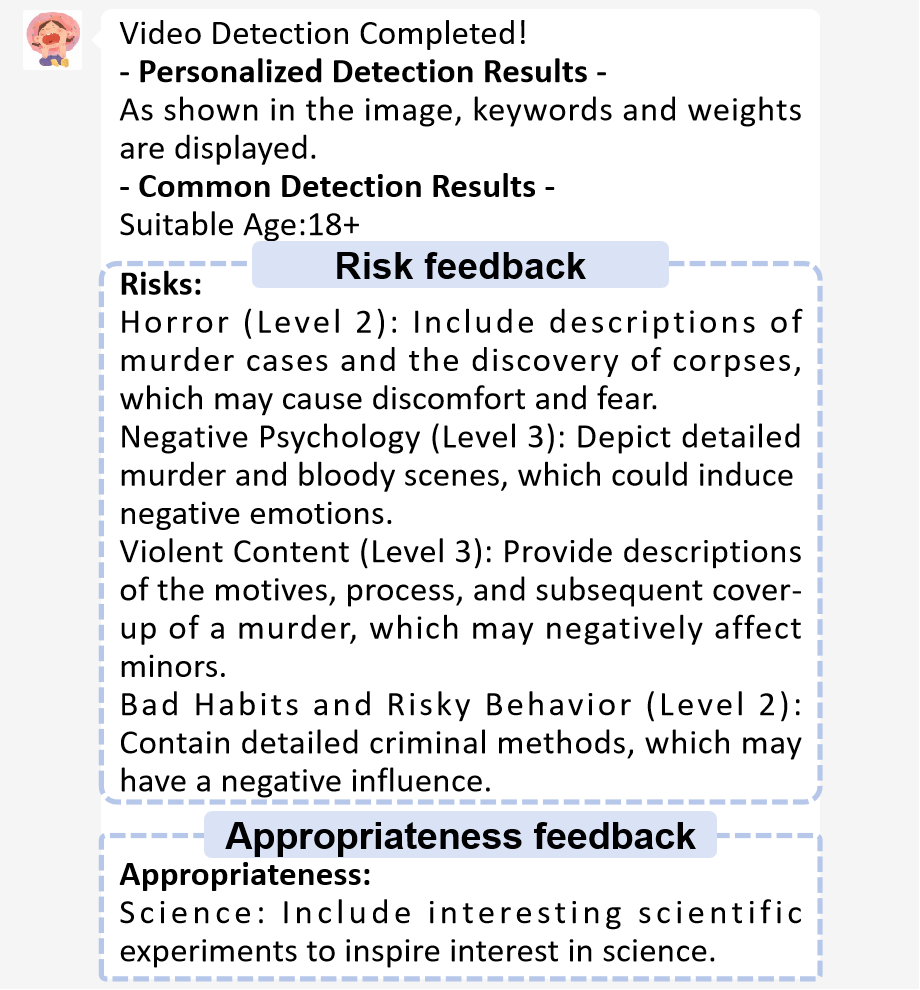}
        \centerline{\small{(b) Common in-time feedback.}}
    \end{minipage}
    \hfill
        \begin{minipage}[t]{0.25\linewidth}
        \centering
        \includegraphics[width=\textwidth]{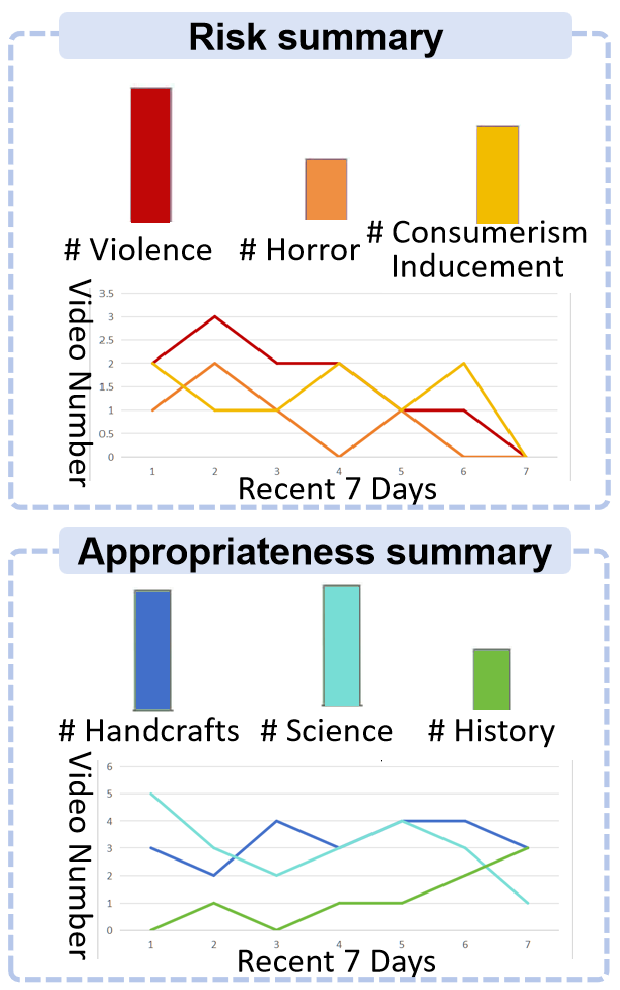}
        \centerline{\small{(c) Common long-term summarization.}}
    \end{minipage}
    \caption{Get feedback and summarization (the interface is translated from Chinese for clarity).}
    \label{fig: in-time feedback}
\end{figure*}

Based on the ``Co-Preference Panel'', the Agent provides both personalized and common in-time feedback. The personalized results, shown in Figure~\ref{fig: in-time feedback} (a), highlight differences between the original co-preference configurations (light-colored bars) and the detection results (dark-colored bars), with matching keywords in green and non-matching ones in red. 
Concurrently, the common detection results provide age-appropriate suitability, risk and appropriateness assessments for videos as Figure~\ref{fig: in-time feedback} (b). 
The summarization process also contains two aspects. The personalized summary, similar to the in-time feedback shown in Figure~\ref{fig: in-time feedback} (a), reflects the average values and the degree of alignment with preference settings over time. 
And the common summary, presented in bar charts and line graphs (Figure~\ref{fig: in-time feedback} (c)), displays the frequency and trends of risk and suitability categories over time. 

\subsection{Summary of YouthCare Characteristics}
To summarize, the characteristics of YouthCare are as follows:

\begin{itemize}
    
    \item \textbf{Collaborative Configuration (D2, D3, D5):} YouthCare enables parents and teenagers to collaboratively configure preferences through both direct and indirect methods via the Collaborative Configuration Module. This module facilitates collaborative configuration through a consensus-building process.

    
    
    \item \textbf{Personalized and Explainable Video Content Censorship (D1, D4):} YouthCare automates video censorship by combining Personalized and Common Guidelines in the Personalized Video Censorship Module and delivering results through the Feedback Report Module as in-time feedback and long-term summarization. This approach provides a solid basis for explanations and helps parents and teenagers understand impacts and discern risks.


    \item \textbf{Modular Design:} YouthCare employs a modular design to ensure that each functional module can be developed and updated independently. This flexibility enables easy adaptation to new requirements and advancements, ensuring the system stays up-to-date and effective.

\end{itemize}


\section{Evaluation}
\label{Evaluation}

\subsection{Settings}


To evaluate the effectiveness of YouthCare, we employed a comprehensive set of metrics, as shown in Table~\ref{tab:Evaluation_Metric}. Drawing from psychological theories and the Technology Acceptance Model (TAM) in the theory of planned behavior \cite{marangunic2015technology}, we organized these metrics into five dimensions to assess each module of YouthCare.

\begin{table*}[ht] 
\small
\caption{Descriptions of evaluation metrics}
\label{tab:Evaluation_Metric}
\centering

\resizebox{\linewidth}{!}{
\begin{tabular}{c|c|c|c|c}
\hline
\textbf{Design Goal} & \textbf{Description}                                                                                                                                     & \textbf{Module}                                                                                                               & \textbf{Metric}              & \textbf{Dimention}            \\ \hline
                     &                                                                                                                                                          &                                                                                                                               & Detection Integrity          & Function                      \\ \cline{4-5} 
                     &                                                                                                                                                          &                                                                                                                               & Detection Accuracy           &                               \\ \cline{4-4}
                     &                                                                                                                                                          &                                                                                                                               & Detection Recall             & \multirow{-2}{*}{Performance} \\ \cline{4-5} 
                     &                                                                                                                                                          &                                                                                                                               & Accessibility                & Design                        \\ \cline{4-5} 
\multirow{-5}{*}{D1} & \multirow{-5}{*}{\begin{tabular}[c]{@{}c@{}}Conduct video censorship \\ from both risk and appropriate perspectives\end{tabular}}                 & \multirow{-5}{*}{Personalized Video Censorship Module}                                                                        & Satisfaction                 & Attitude                      \\ \hline
                     &                                                                                                                                                          &                                                                                                                               & Consensus-Building Integrity & Function                      \\ \cline{4-5} 
                     &                                                                                                                                                          &                                                                                                                               & Preference Understanding     &                               \\ \cline{4-4}
                     &                                                                                                                                                          &                                                                                                                               & Consensus-Building Rate      & \multirow{-2}{*}{Performance} \\ \cline{4-5} 
                     &                                                                                                                                                          &                                                                                                                               & Ease of Use                  & Design                        \\ \cline{4-5} 
\multirow{-5}{*}{D2} & \multirow{-5}{*}{\begin{tabular}[c]{@{}c@{}}Support collaborative configuration of \\ personalized video censorship needs through mediation\end{tabular}} & \multirow{-5}{*}{\begin{tabular}[c]{@{}c@{}}Collaborative Configuration Module\\ (S2: Consensus-Building)\end{tabular}}       & Satisfaction                 & Attitude                      \\ \hline
                     &                                                                                                                                                          &                                                                                                                               & Configuration Integrity      & Function                      \\ \cline{4-5} 
                     &                                                                                                                                                          &                                                                                                                               & Configuration Accuracy       & Performance                   \\ \cline{4-5} 
                     &                                                                                                                                                          &                                                                                                                               & Ease of Use                  & Design                        \\ \cline{4-5} 
\multirow{-4}{*}{D3} & \multirow{-4}{*}{\begin{tabular}[c]{@{}c@{}}Support both direct and indirect methods for \\ configuring personalized detection needs\end{tabular}}            & \multirow{-4}{*}{\begin{tabular}[c]{@{}c@{}}Collaborative Configuration Module\\ (S1: Preference Configuration)\end{tabular}} & Satisfaction                 & Attitude                      \\ \hline
                     &                                                                                                                                                          &                                                                                                                               & Feedback Integrity           & Function                      \\ \cline{4-5} 
                     &                                                                                                                                                          &                                                                                                                               & Feedback Accuracy            &                               \\ \cline{4-4}
                     &                                                                                                                                                          &                                                                                                                               & Personalized Relevance       &                               \\ \cline{4-4}
                     &                                                                                                                                                          &                                                                                                                               & Feedback Acceptance          &                               \\ \cline{4-4}
                     &                                                                                                                                                          &                                                                                                                               & Feedback Guidance Utility    & \multirow{-4}{*}{Performance} \\ \cline{4-5} 
                     &                                                                                                                                                          &                                                                                                                               & Presentation Form            & Design                        \\ \cline{4-5} 
\multirow{-7}{*}{D4} & \multirow{-7}{*}{\begin{tabular}[c]{@{}c@{}}Provide both in-time and long-term explanatory feedback \\ on detection needs\end{tabular}}                  & \multirow{-7}{*}{Feedback Report Module}                                                                                      & Satisfaction                 & Attitude                      \\ \hline
                     &                                                                                                                                                          &                                                                                                                               & Control Integrity   & Function                      \\ \cline{4-5} 
                     &                                                                                                                                                          &                                                                                                                               & Control Frequency           &                               \\ \cline{4-4}
                     &                                                                                                                                                          &                                                                                                                               & Control Effectiveness  & \multirow{-2}{*}{Performance} \\ \cline{4-5} 
                     &                                                                                                                                                          &                                                                                                                               & Controllability              & Design                        \\ \cline{4-5} 
\multirow{-5}{*}{D5} & \multirow{-5}{*}{Ensure user control}                                                                                        & \multirow{-5}{*}{\begin{tabular}[c]{@{}c@{}}All Modules\end{tabular}}                                                                                                  & Satisfaction                 & Attitude                      \\ \hline
\end{tabular}
}
\end{table*}
\begin{itemize}
    \item \textbf{Function:} The integrity of YouthCare is evaluated by examining its functional completeness. This assessment covers the system's ability to offer both direct and indirect configuration options (Configuration Integrity), to support consensus-building between parents and teenagers through the Chatbot (Consensus-Building Integrity), to provide adequate feedback for detection results (Detection Integrity and Feedback Integrity), and to offer active user control participation (Control Integrity).

    \item \textbf{Performance:} The utility of YouthCare is assessed by the performance accuracy. It includes the precision of configuration results (Configuration Accuracy), the Chatbot's ability to comprehend user preferences (Preference Understanding), and the success rate of reaching a consensus (Consensus-Building Rate). Also, it includes how accurately YouthCare detects videos (Detection Accuracy and Detection Recall), and the accuracy, relevance, acceptance and guidance utility of explanations for detection feedback (Feedback Accuracy, Personalized Relevance, Feedback Acceptance and Feedback Guidance Utility). Finally, ``Control Frequency'' and ``Control Effectiveness'' measure the frequency and effectiveness of user engagement in controlling.

    \item \textbf{Design:} The usability of YouthCare is evaluated based on four metrics.
    ``Accessibility'' examines whether the design allows users to readily access the tool when needed. ``Ease of Use'' assesses the user-friendliness of the interface and its operational flow. ``Presentation Form'' focuses on whether the content is presented in an easily understandable manner. Lastly, ``Controllability'' evaluates whether users have adequate control when using YouthCare, enabling them to adjust and customize settings according to their needs.

    \item \textbf{Attitude:} The ``Satisfaction'' metric evaluates users' overall attitudes and impressions toward each module. It explores whether they like the module, have a positive attitude, and feel satisfied with the overall experience.
    
\end{itemize}

In addition to evaluating each module individually, we assess the overall user experience of YouthCare through three metrics: ``Satisfaction'', ``Intention'' and ``Strengths \& Weaknesses''. The ``Satisfaction'' metric measures users' feelings about YouthCare, including its configuration, detection, and feedback components. The ``Intention'' metric examines users' willingness to continue using the system. Finally, the ``Strengths \& Weaknesses'' metric summarizes YouthCare's advantages and disadvantages to identify areas for improvement. All evaluations were conducted using a combination of questionnaires and interviews, with the feedback on strengths and weaknesses primarily relying on interviews.

\begin{table*}[ht] 
\small
\caption{Demographics of evaluation participants.}
\label{tab:Summary of evaluation participants}
\centering
\resizebox{\linewidth}{!}{
\begin{tabular}{c|ccccccccc}
\hline
\textbf{ID}   & \textbf{\begin{tabular}[c]{@{}c@{}}Youth\\ Age\end{tabular}} & \textbf{\begin{tabular}[c]{@{}c@{}}Youth\\ Gender\end{tabular}} & \textbf{\begin{tabular}[c]{@{}c@{}}Parent\\ Age\end{tabular}} &
\textbf{\begin{tabular}[c]{@{}c@{}}Parent\\ Gender\end{tabular}}&
\textbf{\begin{tabular}[c]{@{}c@{}}Parent\\ Education\end{tabular}}&
\textbf{\begin{tabular}[c]{@{}c@{}}Parent\\ Occupation\end{tabular}}&
\textbf{\begin{tabular}[c]{@{}c@{}}Parent Usage\\ (Years/Frequency)\end{tabular}}&
\textbf{\begin{tabular}[c]{@{}c@{}}Youth Usage\\ (Weekday/Weekend)\end{tabular}}&
\textbf{\begin{tabular}[c]{@{}c@{}}Whether Participated\\ in Formative Study\end{tabular}}         \\

\midrule
\textbf{YP1}  & 8                                         & Girl            & 35-44                              & Male      &  \begin{tabular}[c]{@{}c@{}}Bachelor's\\ Degree\end{tabular}  &  \begin{tabular}[c]{@{}c@{}}Company\\ Employee\end{tabular} & \begin{tabular}[c]{@{}c@{}}7 Years or More/\\ Every Day\end{tabular}                             & \begin{tabular}[c]{@{}c@{}}Less than 1 Year/\\ 1-6 Days Per Week\end{tabular}  & Y                                                                    \\ \midrule
\textbf{YP2}  & 9                                         & Girl            & 25-34                              & Female    & \begin{tabular}[c]{@{}c@{}}Associate's\\ Degree or Below\end{tabular}   &  Teacher& \begin{tabular}[c]{@{}c@{}}3-5 Years/\\ Every Day\end{tabular}                             & \begin{tabular}[c]{@{}c@{}}1-3 Years/\\ 1-6 Days Per Week\end{tabular}   & Y                                                                    \\ \midrule
\textbf{YP3}  & 9                                          & Girl                                          & 35-44                                       & Female       & \begin{tabular}[c]{@{}c@{}}Master’s\\ Degree or Above\end{tabular}    &  Teacher&\begin{tabular}[c]{@{}c@{}}5-7 Years/\\ 1-6 Days Per Week\end{tabular}                             & \begin{tabular}[c]{@{}c@{}}1-3 Years/\\ 1-6 Days Per Week\end{tabular}                                  & N                                                                    \\ \midrule
\textbf{YP4}  & 10                                         & Boy                                           & 35-44                                       & Female    & \begin{tabular}[c]{@{}c@{}}Master’s\\ Degree or Above\end{tabular}    &  Teacher&\begin{tabular}[c]{@{}c@{}}5-7 Years/\\ 1-6 Days Per Week\end{tabular}                             & \begin{tabular}[c]{@{}c@{}}1-3 Years/\\ 1-6 Days Per Week\end{tabular}                                     & N                                                                    \\ \midrule
\textbf{YP5}  & 11                                         & Girl                                          & 25-34                                       & Female      &  \begin{tabular}[c]{@{}c@{}}Master’s\\ Degree or Above\end{tabular}    &  Teacher&\begin{tabular}[c]{@{}c@{}}5-7 Years/\\ Every Day\end{tabular}                             & \begin{tabular}[c]{@{}c@{}}3-5 Years/\\ Every Day\end{tabular}                                   & Y                                                                    \\ \midrule
\textbf{YP6}  & 12                                         & Boy                                           & 25-34                                       & Female        & \begin{tabular}[c]{@{}c@{}}Bachelor's\\ Degree\end{tabular}    &  \begin{tabular}[c]{@{}c@{}}Company\\ Employee\end{tabular}&\begin{tabular}[c]{@{}c@{}}3-5 Years/\\ Every Day\end{tabular}                             & \begin{tabular}[c]{@{}c@{}}3-5 Years/\\ 1-6 Days Per Week\end{tabular}                                 & N                                                                    \\ \midrule
\textbf{YP7}  & 13                                         & Boy                                           & 45-54                                       & Female       & \begin{tabular}[c]{@{}c@{}}Associate's\\ Degree or Below\end{tabular}   &  \begin{tabular}[c]{@{}c@{}}Full-time\\ Parent\end{tabular}&\begin{tabular}[c]{@{}c@{}}3-5 Years/\\ Every Day\end{tabular}                             & \begin{tabular}[c]{@{}c@{}}1-3 Years/\\ 1-6 Days Per Week\end{tabular}                                  & Y                                                                    \\ \midrule
\textbf{YP8}  & 14                                         & Boy                                           & 35-44                                       & Female    & \begin{tabular}[c]{@{}c@{}}Master’s\\ Degree or Above\end{tabular}    &  Teacher&\begin{tabular}[c]{@{}c@{}}5-7 Years/\\ 1-6 Days Per Week\end{tabular}                             & \begin{tabular}[c]{@{}c@{}}3-5 Years/\\ Every Day\end{tabular}                                     & N                                                                    \\ \midrule
\textbf{YP9}  & 15                                         & Girl                                          & 25-34                                       & Female     & \begin{tabular}[c]{@{}c@{}}Master’s\\ Degree or Above\end{tabular}    &  \begin{tabular}[c]{@{}c@{}}Full-time\\ Parent\end{tabular}&\begin{tabular}[c]{@{}c@{}}3-5 Years/\\ Every Day\end{tabular}                             & \begin{tabular}[c]{@{}c@{}}3-5 Years/\\ Every Day\end{tabular}                                    & N                                                                    \\ \midrule
\textbf{YP10} & 15                                         & Girl                                          & 25-34                                       & Female      & \begin{tabular}[c]{@{}c@{}}Associate's\\ Degree or Below\end{tabular}    &  Nurse&\begin{tabular}[c]{@{}c@{}}3-5 Years/\\ 1-6 Days Per Week\end{tabular}                             & \begin{tabular}[c]{@{}c@{}}3-5 Years/\\ Every Day\end{tabular}                                   & Y                                                                    \\ \hline

\end{tabular}
}
\end{table*}

Then, we recruited participants to evaluate YouthCare and collected feedback through questionnaires and interviews.

\begin{itemize}
    \item \textbf{Participants:} Besides 5 parent-child pairs from the formative study who expressed willingness to join the evaluation, we also recruited an additional 5 pairs screened by the same criteria outlined in Section~\ref{Participants}. This approach aimed to attract new participants who meet our requirements, thereby enhancing the representativeness and generalizability of our evaluation. Ultimately, a total of 10 pairs participated in the evaluation, as shown in Table~\ref{tab:Summary of evaluation participants}. Each group received a compensation of 50 RMB.

    \item \textbf{Procedure:} Prior to the evaluation, we provided each parent-child pair with a brief overview of the process and obtained their consent for data collection. We then introduced the design goals of YouthCare, demonstrated its functions, and offered instructions to ensure both parents and children could correctly use it. After that, each pair used YouthCare independently on their mobile devices, following three phases as Section~\ref{Example Process}. Pre-configuration was initiated by one party after consultation between the parent and child, with a maximum of 10 minutes allocated for consensus-building. If the time was exceeded, the current results were carried forward to the next phase. During the immediate use process, researchers provided assistance only if participants encountered problems. Finally, they could review the overall feedback on their usage. The entire evaluation lasted between 30 minutes to 1 hour, and all user operations were logged for analysis. Upon completing the evaluation, both parents and teenagers were asked to fill out a 5-point Likert scale questionnaire, and we conducted interviews to gather further insights based on the questionnaire, which were recorded and transcribed. The coding method for the transcript data followed a procedure similar to that described in Section~\ref{formative_procedure}.

     
\end{itemize}

\subsection{Results}
\subsubsection{Observation of User Behaviors}

To understand how parents and teenagers used YouthCare during the evaluation, we analyzed user operation logs and conducted on-site observations. The logs indicated an active use of YouthCare, with an average of 4.45 videos detected per user - 4.8 by parents and 4.1 by teenagers - indicating interest, especially among parents. Based on the observations, we found diverse patterns in how parents and teenagers express preferences, collaborate on adjustments, and respond to system feedback using YouthCare. First, when expressing preferences, most parents and teenagers tended to use indirect methods first and then made direct adjustments. However, some parents and teenagers found it difficult to express preferences clearly, especially for some young teenagers. Also, they showed uncertainty in explaining reasons, and sometimes were unable to provide any when Chatbot asked why they made adjustments. Second, when collaborating with Chatbot in preference settings, both parents and teenagers actively engaged. Teenagers, in particular, usually expressed their preferences first, followed by minor adjustments from parents. Third, the feedback results stimulate parents and teenagers to reflect on and actively adjust their configurations. For example, some of them reconsidered their original preference settings and made adjustments after receiving the detection results.

\subsubsection{Analysis of User Surveys and Interviews}

Then, we conducted a comprehensive evaluation of each module, detailed in Table~\ref{tab:Evaluation_Metric}, to assess effectiveness and user experience through questionnaires and interviews. The 5-point Likert scale questionnaires showed reliability with Cronbach's coefficient alpha \cite{cho2015cronbach} value of $\alpha = 0.7836$ for teenagers and $\alpha = 0.7655$ for parents. The slightly lower $\alpha$  for parents might suggest a broader range of individual needs, but both $\alpha$ values exceed the acceptable threshold of 0.7 \cite{taber2018use}, confirming their reliability.
To enhance comprehension, we reordered the evaluation results based on user experience rather than the design goals.

\begin{figure}[t]
    \begin{minipage}[t]{0.4\linewidth}
        \centering
        \includegraphics[width=\textwidth]{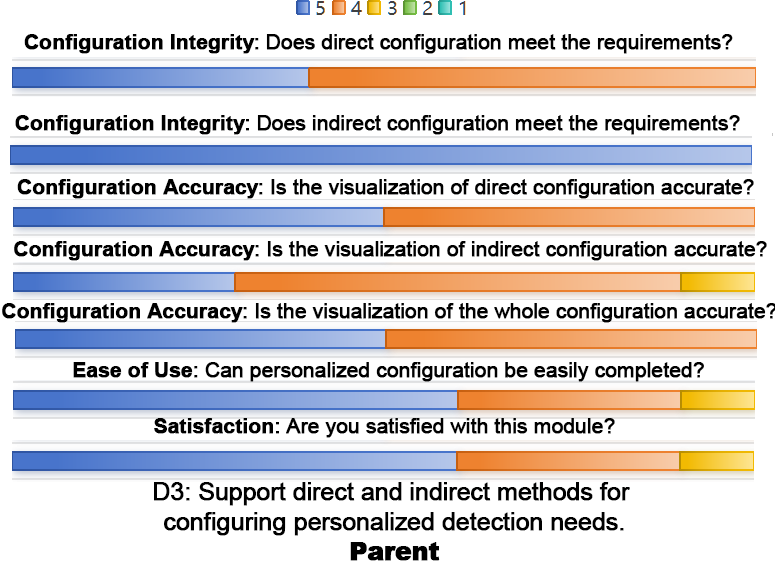}
        \centerline{\small{(a) Statistical results from parents.}}
    \end{minipage}%
    \hfill
    \begin{minipage}[t]{0.4\linewidth}
        \centering
        \includegraphics[width=\textwidth]{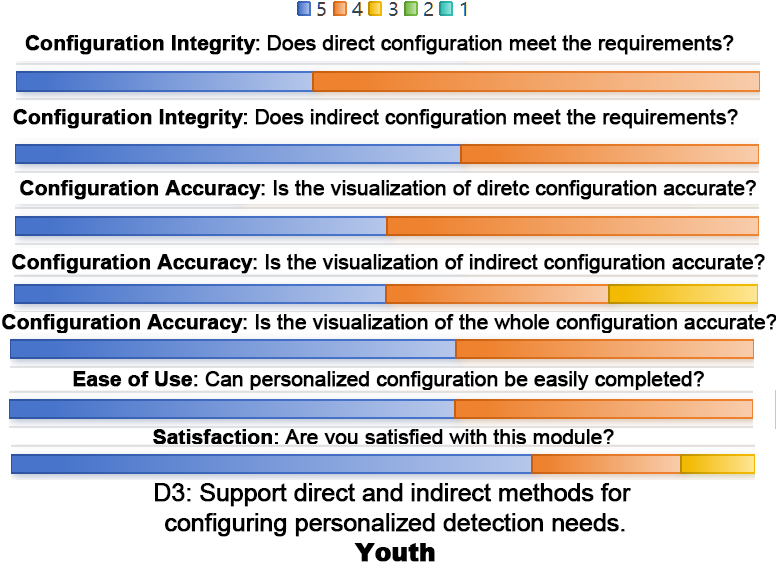}
        \centerline{\small{(b) Statistical results from teenagers.}}
    \end{minipage}
    \vfill
    \caption{Human evaluation results for Personalized Configuration.}
    \label{fig: Eva-D3}
\end{figure}

\textbf{Personalized Configuration (D3):} As shown in Table~\ref{tab:Evaluation_Metric}, this module corresponds to D3 and was evaluated using seven questions (Figure~\ref{fig: Eva-D3}). Both parents and teenagers highly rated YouthCare's personalized configuration feature, with average scores of 4.53 for parents and 4.5 for teenagers.

Specifically, parents found the indirect approach more effective than the direct one, with all rating the indirect method a 5 for configuration integrity, while most parents rated the direct method a 4, mentioning the difficulty in ``articulating my needs directly'' (P3). Teenagers also preferred the indirect configuration, finding it ``useful for expressing preferences directly'' (Y5). Regarding accuracy, half of both groups generally believe direct configuration offers high accuracy, as it ``timely and accurately reflects video preferences'' (Y10). However, most parents and teenagers noted that the indirect method sometimes ``overlooked keywords'' (P2). However, combining both approaches improved the accuracy, with P8 stating it was ``relatively accurate after using both methods''. From the design, more than half of parents and teenagers found the process ``intuitive'' (Y2) and ``convenient'' (Y10), rating it a 5. However, some parents mentioned it ``somewhat lengthy'' (P1) and ``a bit tedious'' (P2). Overall, the personalized configuration feature received positive feedback, with many noting ``quite liked this process'' (P2). 

Overall, the results suggest that the personalized configuration design largely meets user needs, but still needs to be improved in accuracy and convenience.

\begin{figure}[t]
    \begin{minipage}[t]{0.4\linewidth}
        \centering
        \includegraphics[width=\textwidth]{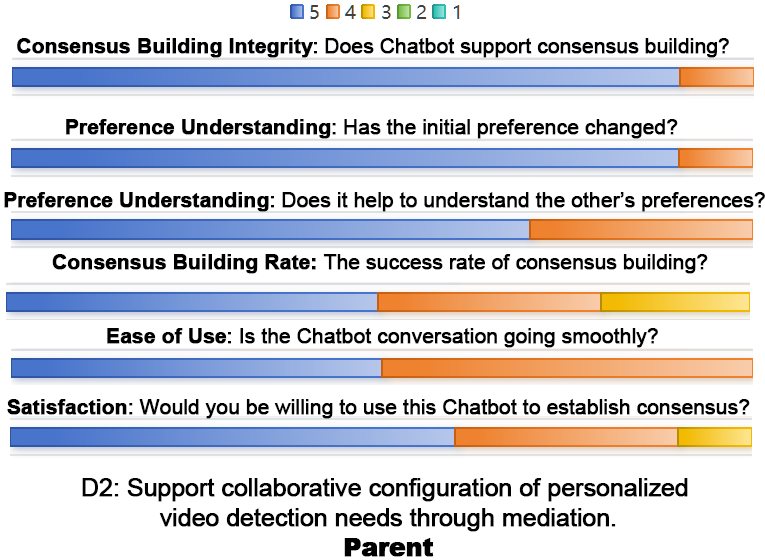}
        \centerline{\small{(a) Statistical results from parents.}}
    \end{minipage}%
    \hfill
    \begin{minipage}[t]{0.4\linewidth}
        \centering
        \includegraphics[width=\textwidth]{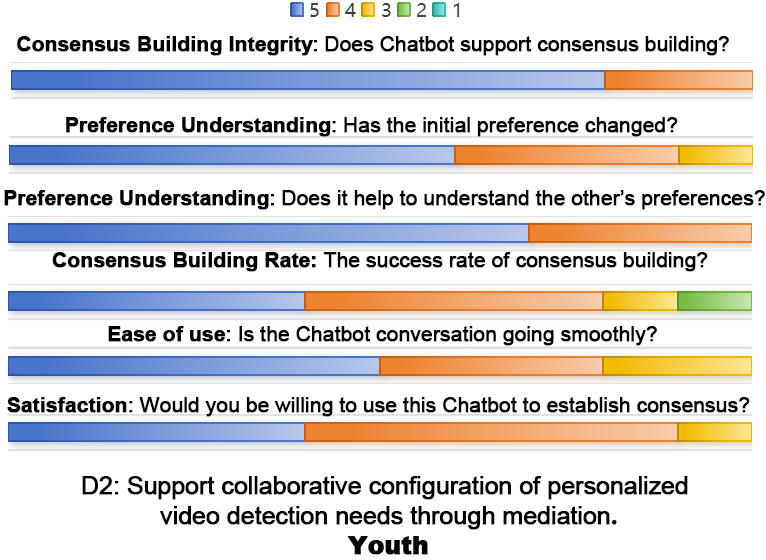}
        \centerline{\small{(b) Statistical results from teenagers.}}
    \end{minipage}
    \vfill
    \caption{Human evaluation results for Collaborative Configuration.}
    \label{fig: Eva-D2}
\end{figure}

\textbf{Collaborative Configuration (D2):} For D2, we introduced a Chatbot to support collaborative configuration between parents and teenagers, aiming to help them reach a consensus. According to user log data, the average number of single-party interactions (i.e., one party sending a message and receiving a reply) was 6.55 rounds, and the average number of mutual interactions (i.e., messages exchanged between parents and teenagers via the Chatbot) was only 3.3 rounds. This indicates a preference for direct communication with the Chatbot rather than exchanges between users. In the collaborative process, almost all pairs ultimately reached a consensus, achieving a success rate of 80\%, while the remaining pairs failed within the given time. For the six questions related to this module, parents had an average score of 4.63, while teenagers averaged 4.4 (Figure~\ref{fig: Eva-D2}). 

Specifically, most parents and teenagers felt that communication with the Chatbot effectively facilitated the consensus-building process, stating ``Chatting on the phone is good; I don't like face-to-face discussions'' (Y5). For the performance, this process helped users modify and understand each other's preferences, as one teenager stated, ``We can understand each other's intentions and thoughts'' (Y10), while a parent noted, ``It allowed me to see clearly what my child likes.'' (P8). However, regarding the consensus-building rate, some parents felt the feature did not fully meet their needs, noting it ``requires some tacit understanding between parents and children'' (P1) and ``takes time to reach a consensus'' (P3). Similarly, most teenagers found consensus-building challenging, with comments like, ``It's hard because there are generational gaps'' (Y9) and ``The success rate is influenced by multiple factors'' (Y10). Additionally, parents rated ease of use and satisfaction slightly higher than teenagers, possibly because parents valued the overall support provided by the functionality, valuing the feature for ``fostering sincere communication between parents and children'' (P1). 

In summary, the collaborative configuration design effectively promoted communication and understanding, but challenges remain in building consensus.

\begin{figure}[t]
    \begin{minipage}[t]{0.4\linewidth}
        \centering
        \includegraphics[width=\textwidth]{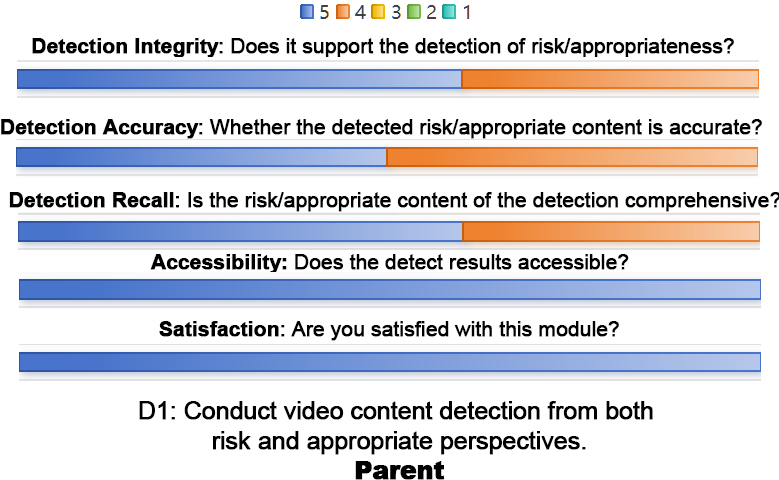}
        \centerline{\small{(a) Statistical results from parents.}}
    \end{minipage}%
    \hfill
    \begin{minipage}[t]{0.4\linewidth}
        \centering
        \includegraphics[width=\textwidth]{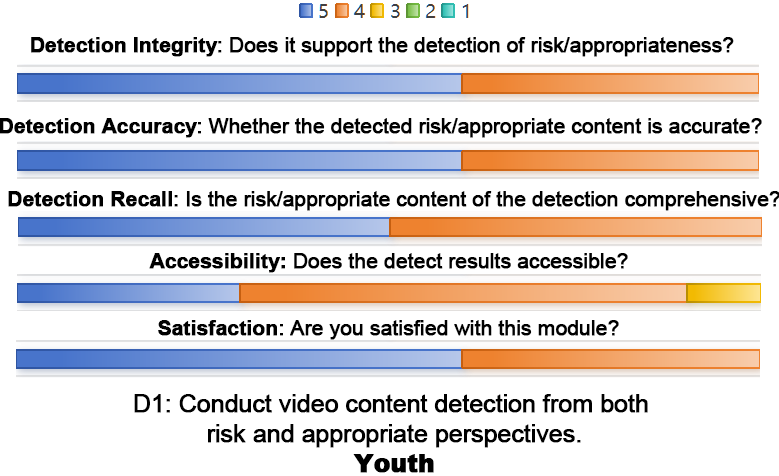}
        \centerline{\small{(b) Statistical results from teenagers.}}
    \end{minipage}
    \vfill
    \caption{Human evaluation results for Video Censorship.}
    \label{fig: Eva-D1}
\end{figure}

\textbf{Video Censorship (D1):} According to D1, YouthCare features a video censorship tool that detects content in terms of risk and appropriateness. This module was assessed through five questions, as shown in Figure~\ref{fig: Eva-D1}. The average scores were 4.74 for parents and 4.46 for teenagers.

Regarding functional support, most parents and teenagers felt their detection needs were met. However, some parents criticized the accuracy of detection results, and some noted missed aspects, such as ``educational elements'' (P9). In contrast, only few teenagers doubted the accuracy, though some of them felt that the explanations were overly sensitive, ``exaggerating minor risks'' (Y8). From the design perspective, while parents rated accessibility highly, teenagers found the video forwarding process ``very inconvenient'' (Y6). Despite these challenges, most users expressed satisfaction with this feature, with one parent stating, ``this function is very necessary and provides substantial support'' (P1). 

These findings suggest that the video censorship module of YouthCare basically meets user needs, but further improvements in detection accuracy, user experience, and interaction process are needed.

\begin{figure}[t]
    \begin{minipage}[t]{0.4\linewidth}
        \centering
        \includegraphics[width=\textwidth]{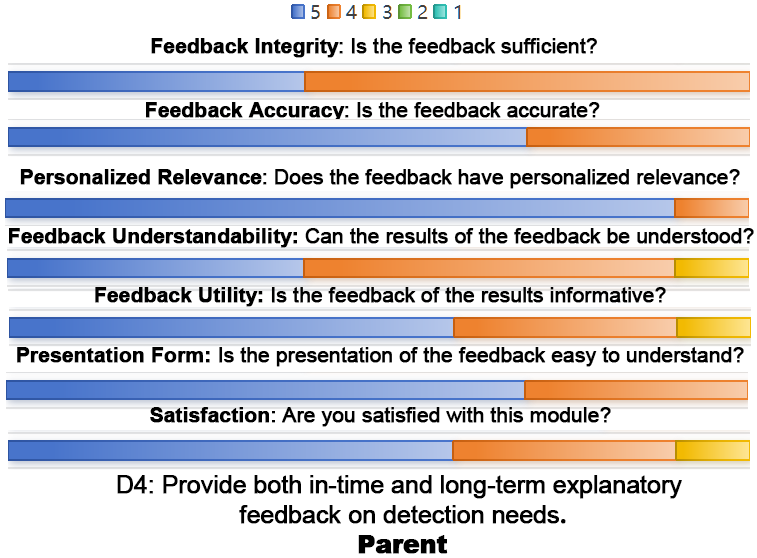}
        \centerline{\small{(a) Statistical results from parents.}}
    \end{minipage}%
    \hfill
    \begin{minipage}[t]{0.4\linewidth}
        \centering
        \includegraphics[width=\textwidth]{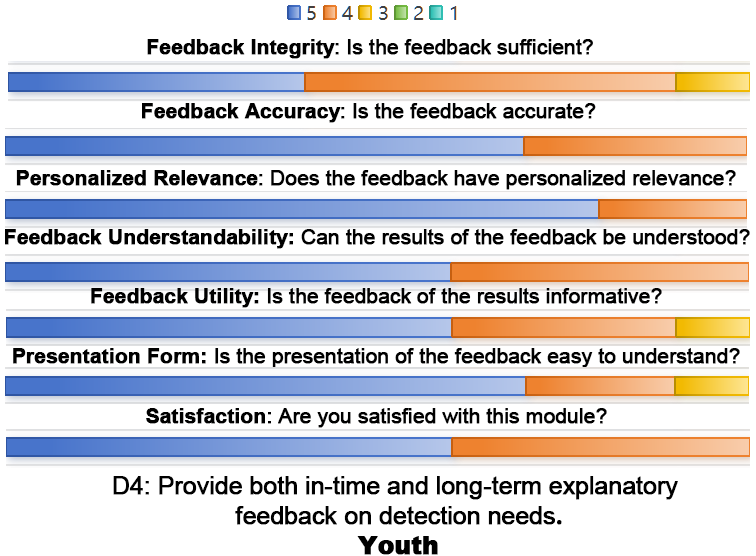}
        \centerline{\small{(b) Statistical results from teenagers.}}
    \end{minipage}
    \vfill
    \caption{Human evaluation results for Feedback and Summarization.}
    \label{fig: Eva-D4}
\end{figure}

\textbf{In-time Feedback and Long-term Summarization (D4):} To enhance the practicality of detection results, YouthCare provides in-time feedback and long-term summarization. Due to limited user engagement, long-term summaries (which summarize the usage process) resemble in-time feedback. This module was evaluated with seven unified questions, as shown in Figure~\ref{fig: Eva-D4}. The results suggest a general consistent rating from parents and teenagers, with average scores of 4.57 and 4.59, respectively.

Regarding functionality, most parents and teenagers found the feedback somewhat insufficient, suggesting ``it could be more detailed'' (P9). Additionally, some parents noted the inaccuracies, such as ``Critical examples were also identified as negative content'' (P1). However, the results were rated highly for their personalized relevance.
Nearly all parents and teenagers acknowledged this high relevance, with comments such as ``the relevance to the given keywords is high'' (P4). However, some of the parents rated feedback understandability as 4, with one rating it 3, noting that ``younger children may lose patience and find it hard to understand'' (P2). Similarly, some teenagers described the content as ``complicated with too much text'' (Y5). Furthermore, most parents recognized its utility in ``helping me make judgments based on real situations'' (P3). Regarding the feedback presentation, most parents and teenagers rated it 5 for its ``clear and easy-to-understand display'' (P5). Moreover, the feedback module was praised for ``providing a detailed understanding of the video in advance'' (Y8) and ``helping me understand my child's situation'' (P4).

The results reveal that the feedback and summarization align well with personalized configurations, which effectively meet users' personalized needs for video censorship and consumption. However, users emphasize the need for clearer and more detailed feedback.




\begin{figure}[t]
    \begin{minipage}[t]{0.4\linewidth}
        \centering
        \includegraphics[width=\textwidth]{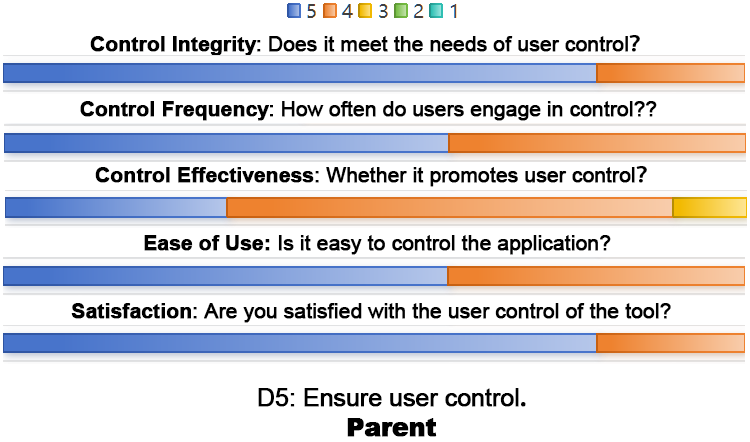}
        \centerline{\small{(a) Statistical results from parents.}}
    \end{minipage}%
    \hfill
    \begin{minipage}[t]{0.4\linewidth}
        \centering
        \includegraphics[width=\textwidth]{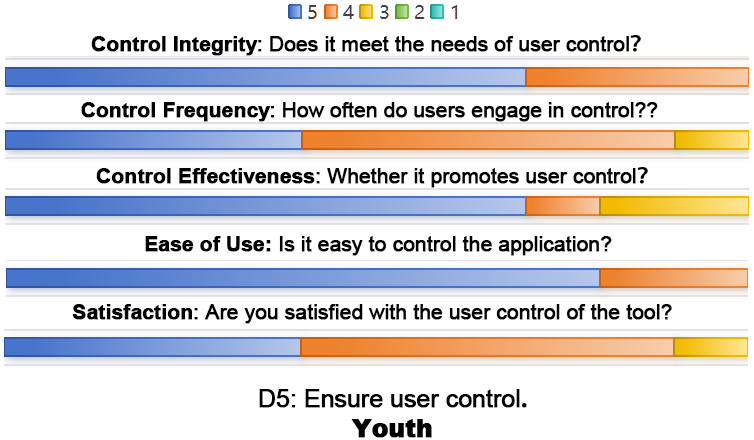}
        \centerline{\small{(b) Statistical results from teenagers.}}
    \end{minipage}
    \vfill
    \caption{Human evaluation results for User Control.}
    \label{fig: Eva-D5}
\end{figure}

\textbf{User Control (D5):} To assess how users can effectively control YouthCare, including direct control over configuration settings and indirect control over the detection and feedback processes, we designed five questions, as shown in Figure~\ref{fig: Eva-D5}. Parents and teenagers gave average scores of 4.6 and 4.42, respectively. 

For its function, most parents and teenagers believed YouthCare fully met their needs for effective control. For example, P2 ``appreciated being involved in my child's usage'', while Y2 ``enjoyed managing my own preference''. When evaluating the frequency of controlling, most parents frequently adjusted the preference panel, as noted by P3, ``I initially adjust preferences, and continuously refine them based on feedback''. In contrast, only a few teenagers made frequent adjustments, with others stating they adjust ``only when feedback does not meet my expectations'' (Y9). Regarding the effectiveness, some parents found ``it is time-saving'' (P7), though others felt ``it lacks efficiency'' (P6). Teenagers also noted that the dialogue contained ``confusing content like obscure keywords'' (Y8), which reduced their engagement. Most parents and teenagers found the control process to be very simple and user-friendly, describing it as ``intuitive and simple to control, allowing quick adjustments of keywords and preferences'' (P1), and ``easy and friendly access, enabling adjustments whenever needed'' (Y6). Moreover, many parents rated the module 5, stating they ``greatly appreciate this involvement in the control process'' (P2), while some teenagers also rated it 5, believing it ``facilitates more convenient and flexible parent-child involvement... offers enough individual freedom'' (Y7).

The results indicate that YouthCare engages parents and teenagers in managing video censorship by allowing intuitive adjustments of keywords and preferences, enabling active and flexible system control. However, improvements in the efficiency and clarity of the feedback process are needed to better support user control.

\subsubsection{Overall}

\begin{figure}[t]
    \begin{minipage}[t]{0.4\linewidth}
        \centering
        \includegraphics[width=\textwidth]{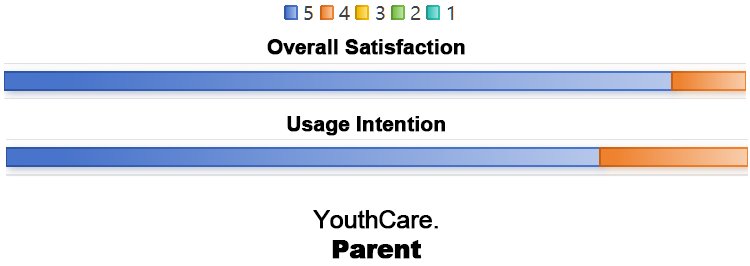}
        \centerline{\small{(a) Statistical results from parents.}}
    \end{minipage}%
    \hfill
    \begin{minipage}[t]{0.4\linewidth}
        \centering
        \includegraphics[width=\textwidth]{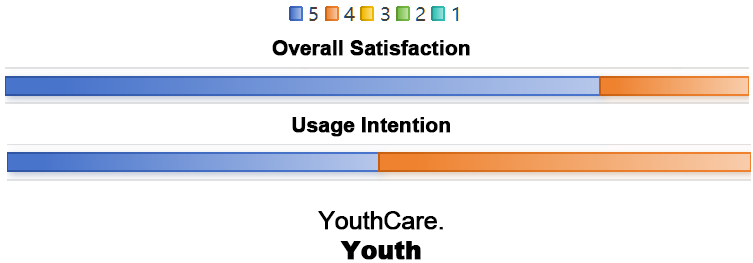}
        \centerline{\small{(b) Statistical results from teenagers.}}
    \end{minipage}
    \vfill
    \caption{Human evaluation results for YouthCare Overall Performance.}
    \label{fig: Eva-all}
\end{figure}

We evaluated YouthCare's overall satisfaction, usage intentions, strengths and weaknesses through additional ratings and interviews (Figure~\ref{fig: Eva-all}). Nearly all parents rated it highly, emphasizing its role in both ``safeguarding teenagers' healthy growth'' (P2) and ``facilitating joint participation'' (P1). Most parents expressed strong intent to continue using it, with one stating, ``especially when concerned about my child's content exposure'' (P8). However, challenges were noted, such as difficulty in ``reaching a consensus'' (P10) and the requirement for ``professional training to improve detection accuracy'' (P3). Among teenagers, most of them gave a rating of 5, stating that YouthCare not only ``helps teens control the content and avoid negative influences'' (Y5) but also ``effectively avoids direct communication issues with parents'' (Y9). Some of the teenagers expressed a strong willingness to continue using it, while others would use it ``selectively'' (Y2). Teenagers were relatively consistent in their opinions on YouthCare's shortcomings, noting that ``the process is somewhat rigid'' (Y3) and ``the different viewpoints may lead to unavoidable conflicts'' (Y10).

This evaluation indicates that YouthCare significantly promotes teenagers' video censorship and parent-child interactions. However, differences in needs and detection accuracy highlight areas for further improvement.

\section{Discussion}
\label{Discussion and Implications}
Drawing from our formative study (Section~\ref{Formative Study}) and evaluation results (Section~\ref{Evaluation}), we obtain several insights into the design of parent-child collaborative video censorship systems.

\subsection{Trade-off of Automated Video Censorship}
With the widespread application of LLMs, video censorship has progressed from rule filtering to deep understanding and reasoning. To achieve a more efficient and accurate comprehension of video content and user preferences, YouthCare introduces MLLMs. Evaluations suggest that although this LLM-powered system improves censorship quality and efficiency, it poses new challenges to human-AI interaction, such as the risk of over-censorship and increased parental dependence.

Firstly, the advanced comprehension capabilities of LLMs greatly enhance the ability of parents and children to censor videos. 
YouthCare helps parents and children accurately identify content that is suitable or unsuitable for children, minimizing inaccuracies such as misjudgments and omissions in content censorship \cite{singh2019kidsguard}. User feedback indicates that parents found the censorship results to be more accurate than expected, and teenagers appreciated the censorship details they had not previously noticed. These advantages significantly boost users' confidence in selecting videos. Secondly, YouthCare enhances the efficiency of video censorship and parent-child interactions. YouthCare adopts automated detection mechanisms that alleviate the burden of manually reviewing each video \cite{zellers2019defending}. 
Moreover, it can offer timely responses and mitigate misunderstandings or conflicts that commonly occur in face-to-face communication between parents and teenagers. Most parents noted that YouthCare's ease of use helps them significantly save time, especially for those with busy work. 
Teenagers also found it easy and quick to select videos of interest using YouthCare. 

However, despite the improvements in accuracy and efficiency brought by automated censorship, issues such as over-censorship and excessive reliance may arise. On the one hand, high-accurate content censorship can sometimes trigger over-censorship, particularly when judging subtle risk content. This may cause parents and teenagers to become overly concerned about the censorship results, and result in unnecessary anxiety about content exposure \cite{brussoni2012risky}. For instance, parents reported that even if a video contains few violent elements, YouthCare generates a significant alert, bringing feelings of anxiety to parents. Future research could focus on refining the degree of content recognition to avoid result exaggeration, e.g., by dynamically adjusting detection sensitivity based on user feedback and real-world scenarios. 
On the other hand, the efficiency of automated detection may lead to parents and children's excessive reliance. Although this system significantly accelerates content filtering and saves time for parents, it may cause them to become overly dependent on it. Parents might overlook their understanding and involvement in children's media use, reducing the direct interaction and communication in the parent-child relationship.
Meanwhile, teenagers' dependence on automated censorship may hinder their development of problem-solving and communication skills \cite{zhai2024effects}. Therefore, how to balance the control boundaries between humans (parents and children) and automated censorship systems to ensure censorship performance while avoiding dependency - becomes a crucial research topic in further video content regulation and family media use. 

\subsection{Facilitating Parent-Child Collaboration in Video Censorship}
Parent-child interactions often face challenges due to the lack of expression of personalized needs and potential conflicts. To address these issues, YouthCare introduces a Chatbot as an intelligent mediator to support parents and teenagers in collaboratively configuring personalized needs during video content censorship. The Chatbot offers opportunities for parents and children to express personalized censorship needs and also gives suggestions for alleviating conflicts, facilitating the efficiency and consensus of parent-child collaboration in video censorship.

The benefits of using Chatbot for parent-child collaboration in video censorship can be summarized as follows. First, Chatbot provides a space for parents and teenagers to express their personal preferences. With advanced natural language processing capabilities, it effectively captures user intentions, ensuring that both parties' needs are clearly expressed and considered. Both parents and teenagers acknowledged that Chatbot effectively helps them express thoughts. Second, Chatbot fosters genuine communication between parents and teenagers by providing a safe mediation environment. This encourages both parties to share their true thoughts, helping build consensus \cite{shin2022chatbots, andersen2021conflicts, beattie2017generating}. Teenagers, in particular, preferred interacting with Chatbot, finding it more free and relaxed than face-to-face communication. Parents also appreciated this indirect approach, as they believed it promotes more sincere dialogue. Additionally, Chatbot enhances mutual understanding and self-awareness. By guiding both parties to share their perspectives and feelings, it helps parents better understand teenagers' interests and needs while also helping teenagers comprehend their parents' educational intentions. Finally, Chatbot improves the comprehensibility and acceptability of communication through role-adapted strategies. For example, it can translate parents' direct communication style into language that is more acceptable for teenagers, thereby balancing the expression of intent and the reception.

While the introduction of Chatbot effectively promotes parent-child collaboration, it also brings some concerns. One concern is the power imbalance stemming from the different roles parents and teenagers play in the family context. The system assumes an equal footing in expressing needs for both parents and teenagers. In reality, teenagers differ from parents in their needs for censorship and their acceptance of feedback, due to the unequal power dynamics inherent in their relationship \cite{omorogiuwa2021power}. Therefore, more flexible communication and feedback mechanisms are necessary to meet the needs of each group. For example, personalized communication strategies could be developed, allowing Chatbot to act as an educational advisor for parents and as a peer-like friend for teenagers, thus enhancing engagement and satisfaction for both parties \cite{wong2019consejero, garg2020conversational}. Another concern is the potential over-reliance on Chatbot, which could reduce opportunities for direct parent-child interactions. While Chatbot is effective in mediating conflicts, its role as an intermediary may lead parents and teenagers to prefer communicating through Chatbot rather than engaging in face-to-face discussions. This dependence might result in parents losing a deeper understanding of teenagers' true thoughts and teenagers relying on the Chatbot's feedback without developing independent judgment skills. Thus, future designs should focus on balancing indirect interaction facilitated by Chatbot with direct communication. For instance, preference expression can be mediated through Chatbot to reduce the pressure and conflict of direct conversation. Parents and teenagers should also be encouraged to engage directly in discussions about values and behavioral norms to foster understanding and build trust. This approach not only improves the efficiency of reaching agreements, but also promotes deeper communication and stronger relationship between parents and teenagers.

\subsection{Impact of Feedback on Family Dynamics}

YouthCare's feedback mechanism has significantly influenced family dynamics by altering communication and shaping relationships between parents and teenagers. However, its use also reveals potential issues, such as privacy breaches and increased family conflicts.

The feedback mechanism in YouthCare has shown significant benefits in enhancing parent-child relationships. By providing detailed feedback, it effectively promotes a shift in dominance from parents to teenagers. According to the Parental Mediation Theory (PMT) framework, parental strategies for influencing children's media use include restrictive mediation, active mediation, and co-using \cite{yu2024parent}. Our study focused on teenagers aged 8 to 15, a critical transition period where media management begins to shift from parent-dominated to teen-dominated. For younger children, parents typically use restrictive mediation to control media use, but as children grow older, they manage media more independently, with parents shifting to active mediation \cite{livingstone2011risks,helsper2013country}. The feedback supports this by offering explanations and summaries of video content risks and appropriateness, helping teenagers develop skills in risk identification, as well as independent judgment and decision-making. Thus, YouthCare promotes the co-using strategy in video censorship through feedback mechanisms, fostering such a shift in dominance within the parent-child relationship. Additionally, the feedback mechanism enhances these relationships by providing detailed reports that help parents better understand teenagers' media preferences and engage in more open and in-depth discussions. 

Despite the benefits of the feedback mechanism in promoting dominance shift within parent-child relationships through in-time feedback and long-term summarization, there are potential drawbacks. First, parents' excessive focus on feedback results may lead to over-scrutiny of their children's media behavior. This heightened attention often exacerbates parental anxiety and may result in tension and conflict with teenagers, as parents might overly interfere with their children's media choices and usage \cite{kanter2012impact}. Second, detailed reports that include teenagers' viewing habits, preferences, and interactions may infringe on their privacy, making them feel constantly monitored. It will diminish their confidence in making decisions and also create feelings of pressure and dissatisfaction, thus affecting their self-expression and openness in communicating with parents. Excessive privacy intrusion complicates family dynamics, as it can hinder the development of teenagers' independence and potentially damage trust and understanding within the parent-child relationship \cite{kanter2012impact}. Future research should focus on optimizing feedback mechanisms to balance their benefits (e.g., fostering dominance shift) and drawbacks (e.g., privacy invasion). For example, developing a more moderated feedback prompt system with thresholds or contextual cues could guide parents in addressing video censorship issues in a more constructive and supportive manner. Additionally, exploring ways to provide effective feedback while respecting teenagers' privacy, such as applying social translucence theory to offer adequate content hints without overly exposing the child's media behavior, could help protect personal information and reduce privacy concerns \cite{niemantsverdriet2016designing,erickson2000social}.

\subsection{Generalizability}


YouthCare is a personalized collaborative video censorship tool demonstrating several advantages in JME. For teenagers, it provides features of risk assessment and suitability analysis of video content, helping to improve their ability to cope with potential risks on video platforms and promote online safety. For parents, it offers various opportunities for them to engage in children's content censorship practices, such as pre-configuring content preferences, conducting preference adjustments, and reviewing summary reports. These features promote parent-child collaboration in JME while emphasizing a balance between parental guidance and teenage autonomy, which makes YouthCare more appropriate for collectivist cultures where parents emphasize content control and families prefer collaborative decision-making (higher socio-economic families, dual-parent families, etc.) in children's media use \cite{horn2013translating, bus2000joint, garg2021understanding, lareau2018unequal}. It provides a consensus-building process, enabling parents and children to set content preferences and adjust viewing strategies collaboratively.


However, in individualistic cultures where children are generally encouraged to make decisions independently and improve self-reliance, or in families with financial or time constraints (low socio-economic families, single-parent families, grandparent-led caregiving families, etc.), parents or grandparents' expectations to involve in children's online content censorship practices can be lower \cite{horn2013translating, bus2000joint, garg2021understanding, lareau2018unequal, rossi2018human}. They tend to rely on straightforward operations (e.g., using predefined settings) or automated features (e.g., platform recommendations) to assist children in making content choices. Under such scenarios, some of YouthCare's collaborative features, like co-configuring content preferences and negotiating agreements, might be seen as tedious and unimportant, limiting YouthCare's general use. For this problem, future improvements should focus on introducing flexible and dynamic participation strategies to cater to the diverse needs of various cultures and families. For example, providing parents with options to flexibly engage in some processes of children's content censorship, e.g., just reviewing and confirming children's settings without co-configuration or accessing feedback without participating in collaborative adjustments. This design would help parents engage in children's media use more flexibly and achieve different levels of engagement. Additionally, dynamic engagement mechanisms can be considered, where the level of parental involvement would be adjusted according to the child's age, risk awareness, and decision-making capabilities. For example, parents would only engage when their child encounters uncertainty or needs help, ensuring the different features of the system can be utilized dynamically and appropriately according to the needs of different families.

\section{Ethical Considerations}
\label{Ethical Consideration}

In developing YouthCare, we prioritized ethical considerations, especially for teenagers, by implementing privacy protections like data anonymization and encryption and adhering to privacy regulations. We ensured transparency in the system’s functions to build trust and accountability. For teenagers, we followed ethical guidelines by obtaining consent from parents (or guardians) before use. We also embedded mechanisms to protect their autonomy by allowing independent use with alerts for risky content. This ensures protection against undue influence in video detection. The system was designed to balance and respect both parents' and teenagers' perspectives, facilitating collaborative consensus rather than exerting control. Through these efforts, YouthCare aims to be a safe, respectful, and supportive tool for enhancing family interactions and guiding healthy media consumption among teenagers.

\section{Limitations and Future Works}
\label{Limitations and Future Works}
As an exploratory study on designing parent-child collaborative video censorship, this work has the following limitations.

First, younger teenagers may struggle with YouthCare's text-based interface, resulting in inadequate interaction and difficulty in expressing preferences accurately during configuration and feedback processes. This limitation may also lead to misinterpretations or information loss in censorship. Future efforts could focus on improving interaction design by incorporating multimodal methods, such as voice input and graphical interfaces, to better support the needs of teenage users. In addition, involving more teenagers in system design and testing can help align detection methods with their usage habits and cognitive levels.

Second, the evaluation mainly focused on short-term effects and did not explore long-term impacts over extended periods, such as one week or more. Conducting long-term evaluation is essential to understand the system's sustained effects and identify potential issues. Future work could involve extended evaluations to collect data on the system's enduring impact on families and teenagers, with regular follow-ups to gather user feedback and usage patterns, ensuring continuous improvement of the system. 

Lastly, while LLMs exhibited strong analytical and reasoning capabilities with multimodal content, they may lack sensitivity to the specific needs and contexts of parent-child scenarios. Additionally, the reliance on pre-trained data may introduce biases or inaccuracies, especially with diverse cultural backgrounds or specific family communication. Furthermore, LLMs may struggle with complex video content, such as distracting visuals or frequent scene changes, leading to potential misinterpretations and biased detection results. Future research should focus on adapting and training LLMs for family interactions, using customized data and technical enhancements to improve accuracy and reduce biases, and ensuring fairness and reliability across various user groups.

\section{Conclusion}
\label{Conclusion}

In this work, we aimed to address the limitations of traditional parent-child Joint Media Engagement (JME) mechanisms by designing YouthCare, a personalized collaborative video censorship tool to facilitate more efficient and effective video censorship in JME. Through a formative study, we identified parents' and teenagers' needs and expectations regarding personalization and collaboration, which informed the design of YouthCare to support the filtering and selection of video content. An evaluation with 10 parent-child pairs demonstrated YouthCare's effectiveness in enhancing content control and collaboration while also identifying areas for improvement, such as addressing the differing needs of parents and teenagers and enhancing the efficiency and effectiveness of the feedback process. These findings suggest that future collaborative parent-child JME system could consider how to balance control boundaries between human and AI, indirect and direct communications, and dominance and privacy.


\begin{acks}  
This work is supported by National Natural Science Foundation of China (NSFC) under the Grant No. 62372113 and 61932007. Tun Lu is also a faculty of Shanghai Key Laboratory of Data Science, Fudan Institute on Aging, MOE Laboratory for National Development and Intelligent Governance, and Shanghai Institute of Intelligent Electronics \& Systems, Fudan University.
\end{acks}

\bibliographystyle{ACM-Reference-Format}
\bibliography{sample-base}

\appendix

\section{Appendix}
\subsection{YouthCare's Prompts}
\label{Appendix_Prompts}

\begin{figure}[t]
    \begin{minipage}[t]{0.4\linewidth}
        \centering
        \includegraphics[width=\textwidth]{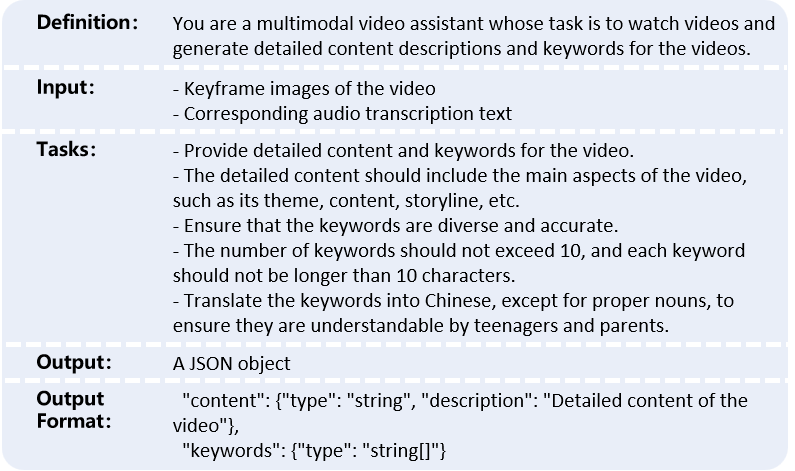}
        \centerline{\small{(a) Prompt for single video feature extraction.}}
    \end{minipage}%
    \hfill
    \begin{minipage}[t]{0.4\linewidth}
        \centering
        \includegraphics[width=\textwidth]{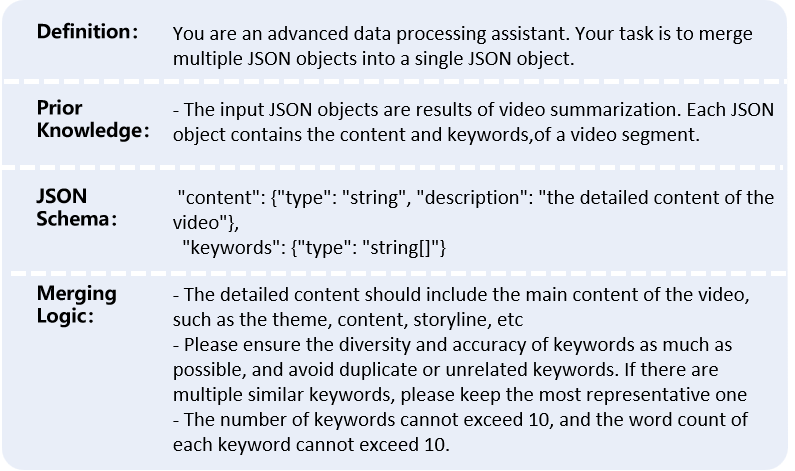}
        \centerline{\small{(b) Prompt for video segment combination.}}
    \end{minipage}
    \vfill
    \caption{Prompts for video feature extraction.}
    \label{fig: appendix_feature_promts}
\end{figure}

The video feature extraction prompt is illustrated in Figure~\ref{fig: appendix_feature_promts} (a). For longer videos exceeding GPT-4o's context length, we divided them into segments and processed each segment through the workflow as shown in Figure~\ref{fig: appendix_feature_promts} (b) to generate JSON strings, which were then combined. The video censorship process follows a similar approach using different output formats per design.


\end{document}